\documentclass[12pt]{article}
\usepackage{authblk}
\usepackage{amsmath}
\usepackage{graphicx}     %
\usepackage{psfrag,epsf}
\usepackage{enumerate}
\usepackage{natbib}
\usepackage{url} 
\usepackage{fullpage}
\usepackage{verbatim}
\usepackage{setspace}
\usepackage{amsbsy}
\usepackage{amsthm}
\usepackage{amssymb}
\usepackage{amsmath}
\numberwithin{equation}{section}
\newtheorem{theorem}{Theorem}[section]
\newtheorem{lemma}[theorem]{Lemma}
\newtheorem{proposition}[theorem]{Proposition}
\newtheorem{corollary}[theorem]{Corollary}
\usepackage[capposition=top]{floatrow}
\usepackage{caption}
\usepackage{mathrsfs}
\usepackage{multirow}
\usepackage{subfigure} 
\usepackage{mathtools}
\usepackage{makecell}
\usepackage[ruled,vlined]{algorithm2e}
\usepackage{longtable}
\usepackage{floatrow}
\usepackage{varwidth}
\usepackage{multirow}

\newcommand{\bel}{\begin{eqnarray}\label}
\newcommand{\eel}{\end{eqnarray}}
\newcommand{\bes}{\begin{eqnarray*}}
	\newcommand{\ees}{\end{eqnarray*}}
\let\bar\overline

\usepackage{xcolor}
\usepackage[colorlinks=true,
linkcolor=blue,
urlcolor=blue,
citecolor=blue,
hypertexnames=false]{hyperref}

\def\cov{{\rm Cov}}
\def\var{{\rm Var}}

\let \tilde \widetilde
\let \hat \widehat

\theoremstyle{definition}
\newtheorem{definition}{Definition}[section]

\allowdisplaybreaks

\begin{document}
	\setlength{\abovedisplayskip}{5pt}
	\setlength{\belowdisplayskip}{5pt}
	\setlength{\abovedisplayshortskip}{5pt}
	\setlength{\belowdisplayshortskip}{5pt}
	
	\title{\LARGE Robust Functional Principal Component Analysis via Functional Pairwise Spatial Signs} 
	\author[a]{Guangxing Wang}
    \author[b]{Sisheng Liu}
    \author[c]{Fang Han}
    \author[a]{Chongzhi Di}
    \affil[a]{Division of Public Health Sciences, Fred Hutchinson Cancer Research Center, Seattle, WA, USA}
    \affil[b]{Kuaishou Technology Co., Beijing, China}
    \affil[c]{Department of Statistics, University of Washington, Seattle, WA, USA}
	\date{December 15, 2020}
	\maketitle
	\begin{abstract}
		Functional principal component analysis (FPCA) has been widely used to capture major modes of variation and reduce dimensions in functional data analysis. However, standard FPCA based on the sample covariance estimator does not work well in the presence of outliers. To address this challenge, a new robust functional principal component analysis approach based on the functional pairwise spatial sign (PASS) operator, termed PASS FPCA, is introduced where we propose estimation procedures for both eigenfunctions and eigenvalues with and without measurement error. Compared to existing robust FPCA methods, the proposed one requires weaker distributional assumptions to conserve the eigenspace of the covariance function. In particular, a class of distributions called the weakly functional coordinate symmetric (weakly FCS) is introduced that allows for severe asymmetry and is strictly larger than the functional elliptical distribution class, the latter of which has been well used in the robust statistics literature. The robustness of the PASS FPCA is demonstrated via simulation studies and analyses of accelerometry data from a large-scale epidemiological study of physical activity on older women that partly motivates this work.
	\end{abstract}
	\noindent%
	\vfill

\section{Introduction} \label{Intro}

Functional principal component analysis (FPCA) is an important tool for capturing the major modes of variation in functional and longitudinal data across many scientific areas, including public health, genetics, finance, and social sciences. An overview of FPCA and its variations can be found in \cite{ramsay2006functional} and \cite{Wang2015}. \citet*{dauxois1982asymptotic} and \cite{hall2006properties}, among many others, studied various aspects of FPCA when functions or curves are fully observed. FPCA for densely recorded sample curves has been investigated by for example \citet*{castro1986principal} and \cite{cardot2000nonparametric}. 
For functional data recorded on a set of sparse and irregular grid points, the FPCA approach was studied by, for instance,  \cite{staniswalis1998nonparametric} and \cite{yao2005functional}. However, the methods mentioned above mostly rely on sample covariance estimators that are sensitive to heavy-tailedness and outliers; therefore, in many applications, methods that are robust to these data challenges are desired. 

Existing works on robust FPCA have already shown promising results. To name a few, \cite{Locantore1999} and \cite{Gervini2008} studied the spherical principal component approach which will be contrasted to ours; \cite{Bali2011} extended the projection-pursuit approach, which replaces the variance with a robust scale estimator in the maximization problem, from multivariate data to function data; \cite{kraus2012dispersion} proposed to replace the covariance operator by a dispersion operator, which is an M-estimator of the covariance operator; \cite{boente2015s} took a slightly different route by exploiting the fact that the sample eigenfuctions provide a subspace that empirically best approximates the original data, and replaced the residuals in the minimization process with robust scale estimators. 

All aforementioned approaches, however, require a certain notion of symmetry --- e.g, an elliptically shaped density contour --- on the studied data. This could be too restrictive in many scientific applications as can be seen in the following motivating example, where the data are highly skewed. Consider the Objective Physical Activity and Cardiovascular Disease Health (OPACH) study \citep{LaCroix2017}, which used accelerometer to objectively measure physical activity in over 6,500 older women. Accelerometers are wearable devices that continuously record a person's movement in three dimensions, and in the OPACH study, the raw data resolution is 30 Hertz. The acceleometry data were then summarized into 1-minute epochs by an index called activity counts, so that each day is a time series of 1,440 data points. The goal of this study is to investigate the association between physical activity patterns and cardiovascular health among older women, and the data on 6,500 women are a typical example of densely recorded functional data. Figure \ref{sec app: acc data boxplots} in Section \ref{sec application} provides an example of the OPACH data. It is evident that at each time point the distribution is skewed and that outliers may exist. This data thus call for robust methods yet need one that is able to waive symmetry assumptions on it.

Partly motivated by the accelerometry data from the OPACH study, we propose a robust FPCA method that answers the aforementioned call. Our method is based on the pairwise spatial sign (PASS) covariance function to be introduced in Section \ref{sec meth} (cf. Equations \ref{sec meth eq: pass} and \ref{sec meth eq: sample psscf}). There the enforced pairwise difference implementation provably waives symmetry requirements by construction, while the spatial sign structure ensures that the influence of an outlier is tamed down. The intuition for the robustness is that the pairwise spatial sign covariance intentionally ignores the information about how far each data unit is from the data center, and instead relies solely on the data's relative direction to each other; thus, an outlier that is far from the data center will not have a severe adverse effect on the sample pairwise spatial sign covariance matrix. 

A natural question is whether the estimates based on the sparial sign covariance are interpretable. Surprisingly, under merely a coordinatewise symmetry condition that is strictly weaker than the elliptical \citep[cf. Equation 2 in][]{Marden1999},  \cite{Marden1999} showed that the population pairwise spatial sign covariance matrix shares the population covariance matrix's eigenspace, thus suggesting the use of the sample pairwise spatial sign covariance matrix in place of the sample covariance matrix to provide robust eigenvector estimation. For a review of the pairwise spatial sign covariance matrix, we refer to \citet*{Taskinen2012}, \citet{han2017eca}, and the references therein. 

This paper studies the FPCA method based on the PASS covariance function under a fully functional setting. In contrast to \citet{Marden1999}, \cite{Locantore1999}, and \cite{Gervini2008}, our new insight appears to be on the pressing importance of conducting pairwise difference before taking self-normalization, which allows for further robustness to asymmetric data --- a phenomenon that is apparently less known in literature. In addition, besides the  estimation of eigenfunctions, we further study the problem of eigenvalue estimation, a topic that has not been studied much in the existing robust FPCA literature. To this end, we exploit a conservative signal preservation pattern in comparing eigenvalues of the PASS covariance function to those of the regular covariance function (cf. Corollary \ref{sec meth cor: eigen ratio relation} in Section \ref{sec meth}) and we further present an  algorithm to approximate their exact values; this algorithm is to be shown in Section \ref{subsection: Eigenvalue estimation}.

The rest of this paper is organized as follows. Section \ref{sec meth} introduces the robust FPCA method applied to data without measure errors. This section is separated to three subsections. Section \ref{subsection: background and notation} lays out the background and notations. Section \ref{subsection: Robust FPCA} introduces the definition of the newly introduced weakly FCS distribution family and the correspondingly proposed robust FPCA method. Section \ref{sec meth subsection: theory of PASS} introduces theoretical properties of the proposed method with relevant proofs relegated to the Appendix A. Section \ref{subsection: Eigenvalue estimation} introduces a robust eigenvalue estimation procedure based on the proposed robust FPCA method. Extensions to functional data measured with noise are presented in Section \ref{subsection; extension to measurement with noise}. Section \ref{sec simulation} contains simulation studies. Application to the OPACH data is presented in Section \ref{sec application}. 

\section{Noise-free Robust FPCA} \label{sec meth}

\subsection{Background and notations} \label{subsection: background and notation}

Consider a random process $X(\cdot)$ in a square integrable Hilbert space $L^2(I)$ over a compact set $I \subset \mathbb{R}$, where the inner product is defined as $\langle f,\ g\rangle := \int_I f(t)g(t) dt \text{ for } f,g \in L^2(I)$ with the induced norm $\|f\| := (\langle f,\ f\rangle)^{1/2} < \infty$. Without loss of generality, assume $ I = [0,1] $ in the rest of this paper. Let $\mu(\cdot)$ be the mean function of $X(\cdot)$ and 
\begin{align} \label{sec meth eq: cov fun}
    \Gamma(s,t) := \cov \{X(s), X(t)\},\ s, t \in [0,1]
\end{align}
be the covariance function of $X(\cdot)$. Assume further that $\Gamma(\cdot,\cdot)$ is continuous and positive definite. We define the covariance operator as $\mathbf{\Gamma} (f)(\cdot) := \int_I \Gamma(s, \cdot) f(s) ds$.

By Mercer's Lemma \citep[cf. Lemma 1.3 in][]{Bosq2012}, the covariance function \eqref{sec meth eq: cov fun} can be decomposed as $\Gamma(s,t) = \sum_{j = 1}^\infty \lambda_j(\Gamma) \phi_j(s) \phi_j(t)$, where $\phi_j(\cdot),\ j = 1, 2, \dotsc$ are orthonormal eigenfunctions of $\mathbf{\Gamma}$ and their corresponding eigenvalues $\lambda_j(\Gamma),\ j = 1, 2 \dotsc$ are in non-increasing order, $\lambda_1(\Gamma) \geq \lambda_2(\Gamma) \geq \cdots \geq 0$. Furthermore, by Karhunen-Lo\`eve expansion \citep[cf. Theorem 1.5 in][]{Bosq2012}, $X(\cdot)$ can be expressed as
\begin{align} \label{sec meth eq: KL expansion}
    X(\cdot) = \mu(\cdot) + \sum_{j = 1}^\infty \xi_j \phi_j(\cdot),
\end{align}
where $\xi_j := \langle X - \mu,\ \phi_j\rangle,\ j = 1, 2, \dotsc$ are the scores of $X(\cdot)$ and they satisfy the properties $\mathbb{E}(\xi_j) = 0 ~\text{ and }~ \cov(\xi_i, \xi_j) = \lambda_j(\Gamma) 1(i=j)$, where $1(\cdot)$ represents the indicator function.

The expansion \eqref{sec meth eq: KL expansion} provides an important tool for dimension reduction in functional data analysis and its performance depends on how well the eigencomponents $\lambda_j(\Gamma)$ and $\phi_j(\cdot)$ are able to be estimated from the sample curves $\Big\{x_1(\cdot), ~x_2(\cdot), ~\ldots, ~x_n(\cdot)\Big\}$.

Traditionally, $\lambda_j(\Gamma)$'s and $\phi_j(\cdot)$'s can be estimated respectively by $\lambda_j(\hat{\Gamma})$ and $\hat{\phi}_j(\cdot)$, which are the eigencomponents of the sample covariance function 
\begin{align} \label{sec meth eq: sample cov fun}
    \hat{\Gamma}(s,t) := \frac{1}{n-1} \sum_{j=1}^n \Big\{x_j(s) - \bar{x}(s) \Big\}\Big\{ x_j(t) - \bar{x}(t)\Big\},~~~ s, t \in [0,1]
\end{align}
with $\bar{x}(\cdot) := n^{-1} \sum_{j=1}^n x_j(\cdot)$ representing the sample mean function. However, the sample covariance estimators do not perform well when the data are heavy-tailed or are contaminated by outliers. As a result, in many applications more robust FPCA methods are desired.

\subsection{PASS FPCA} \label{subsection: Robust FPCA}

To introduce the proposed robust estimators of the eigencomponents, we first define the population PASS covariance function as
\begin{align} \label{sec meth eq: pass}
    K(s,t) = \mathbb{E}\left[ \frac{\{X(s)-\widetilde{X}(s)\} \{X(t)-\widetilde{X}(t)\}}{\| X(\cdot)-\widetilde X(\cdot) \|^2} \right], \text{ for any } s,t\in[0,1],
\end{align}
where $\widetilde{X}(\cdot)$ is an independent copy of $X(\cdot)$. The PASS covariance function is the regular covariance function of the pairwise spatial sign $\{X(\cdot)-\tilde X(\cdot)\}/\|X(\cdot)-\tilde X(\cdot)\|$, thus earning its name. In practice, $K(\cdot,\cdot)$ can be estimated using the following sample PASS covariance function, a standard U-statistic by construction:
\begin{align} \label{sec meth eq: sample psscf}
    \hat{K}(s,t) := \frac{2}{n(n-1)} \sum_{1 \leq j < k \leq n} \frac{ \Big\{x_j(s) - x_k(s)\Big\} \Big\{x_j(t) - x_k(t)\Big\} }{ \Big\| x_j(\cdot) - x_k(\cdot) \Big\|^2 }, \text{ for any } s,t\in[0,1].
\end{align}

In contrast to the sample covariance function, the sample PASS covariance function $\hat{K}(\cdot,\cdot)$ mitigates the part where the sample covariance function $\hat{\Gamma}(\cdot,\cdot)$ is susceptible to outliers. The pairwise difference construction in \eqref{sec meth eq: sample psscf} avoids the explicit calculation of the sample mean $\bar{x}(.)$, which could be heavily influenced by outliers. More importantly, the projection of the differences onto the unit sphere eliminates the magnitude information which is exactly where an outlier heavily exerts its distortion on $\hat{\Gamma}(\cdot,\cdot)$. 

In principal component analysis, it is known that eigenfunctions represent the directions of variation and their corresponding eigenvalues measure the sizes of variation in their respective directions. Thus, the sacrifice of magnitude information should not affect the eigenfunctions. This intuition will be  confirmed by a subsequent theorem, Theorem \ref{sec meth thm: Functional Oja}, where we will prove that under some regularity conditions the eigenfunctions of $K(\cdot,\cdot)$ and $\Gamma(\cdot,\cdot)$ are identical. To achieve full rigor in establishing this claim, however, we have to first regulate the data distribution; these shall occupy the rest of this subsection.

The elliptical distribution class plays a pivotal role in multivariate robust statistics literature. It is a direct relaxation of the Gaussian distribution, able to capture heavy-tailedness and nontrivial tail dependence between variables. A class of multivariate distributions that is larger than the elliptical was introduced in \cite{Marden1999}. These distribution classes can be generalized to functional settings and they also play important roles in functional robust statistics. In \citet{Bali2009}, the authors extended the elliptical distribution class to the functional space; we first briefly sketch their definition of the functional elliptical distribution, then we introduce our extension of the distribution introduced in \cite{Marden1999}. We include the multivariate elliptical distribution and the model in \cite{Marden1999} in the Appendix A.

\begin{definition}[\citeauthor{Bali2009}, \citeyear{Bali2009}] \label{sec meth def: functional elliptical distribution}
   A random element $X(\cdot)$ in a separable Hilbert space $\mathcal{H}$ is said to follow a \emph{functional elliptical} (FE)  distribution with parameters $\mu(\cdot) \in \mathcal{H}$ and $\mathbf{\Gamma}_X: \mathcal{H} \rightarrow \mathcal{H}$ as a self-adjoint, positive semi-definite, and compact operator if and only if for any integer $d \geq 1$ and any linear and bounded operator $A: \mathcal{H} \to \mathbb{R}^d$, we have $AX \sim \epsilon_d(A\mu, A \mathbf{\Gamma}_X A^*, \varphi)$, where $A^* : \mathbb{R}^d \to \mathcal{H}$ denotes the adjoint operator of $A$ and $ \varphi$ is the characteristic generator. In this case we write $X(\cdot) \sim E(\mu(\cdot), \mathbf{\Gamma}_X, \varphi)$, and call $ \mu(\cdot) $ the \emph{location function} and $ \mathbf{\Gamma}_{X} $ the \emph{scatter operator} of $ X(\cdot) $.
\end{definition}

FE distribution is a strict extension to the multivariate elliptical distribution and reduces to it when $\mathcal{H}$ is a regular real space coupled with the Euclidean norm. This indicates that FE distribution, similar to the classic elliptical distribution, encompasses a promising way to model heavy-tailed datasets. On the other hand, like the elliptical distribution, FE distribution requires much on data symmetry, a condition too restrictive to hold in many applications. Motivated by this, we introduce the main distribution class this paper is focused on; as shall be seen soon, it strictly extends the FE class and, in particular, is able to fully eliminate the symmetry requirement. 

\begin{definition}[Weakly FCS distribution class] \label{sec meth def: functional coordinate symmetric}
A random element $X(\cdot)$ in a separable Hilbert space $ \mathcal{H} $ with mean $ \mu(\cdot) $ is said to follow a \emph{functional coordinate symmetric} (FCS) distribution if and only if for any positive integers $ d \leq d' \leq \infty$ with $d<\infty$ and any othonormal bases $\{\psi_1(\cdot), \ldots, \psi_d(\cdot)\}$ in $\mathcal{H}$, 
we have  
\[
\Big(\langle X-\mu, \psi_{1} \rangle, \dotsc, \langle X-\mu, \psi_{d} \rangle\Big)^{\top} \overset{D}{=} \Omega_{\psi} Z_{\psi},
\]
where
	$ \Omega_{\psi} $ is a $ d \times d' $ matrix only depending on $\{\psi_1(\cdot),\ldots,\psi_d(\cdot)\}$ such that $ \Omega_{\psi} \Omega_{\psi}^{\top} = I_d $, the $d$-dimensional identity matrix, and $ Z_{\psi} \in \mathbb{R}^{d'}$ is coordinatewise symmetric. Furthermore, the random element $ X(\cdot) $ is \emph{weakly functional coordinate symmetric} (wFCS) if $ X(\cdot) - \tilde{X}(\cdot) $ is FCS, where $ \tilde{X}(\cdot) $ is an independent copy of $ X(\cdot) $.
\end{definition}
The following proposition shows that, as the FE distribution is an extension of the multivariate elliptical distribution, the FCS distribution class extends the model considered in \cite{Marden1999}.
\begin{proposition} \label{sec meth: FCS equivalent def}
The FCS reduces to the model in \citet{Marden1999} when the separable Hilbert space $ \mathcal{H} $ is an Euclidean space coupled with the regular inner product.
\end{proposition}
The next proposition shows that the wFCS distribution class, which is the main working model in this paper, is strictly larger than the FCS distribution class that is further strictly larger than the FE one.
\begin{proposition} \label{sec meth: relations between functional distributions}
The following relation holds:
	\begin{align*}
		 \text{FE distribution class} ~\subset~ \text{FCS distribution class} ~\subset~ \text{wFCS distribution class}.
	\end{align*}
\end{proposition}
Our last result in this subsection shows that the wFCS distribution allows for arbitrary margins. 
\begin{proposition} \label{sec meth: no marginal distribution required}
For any well-defined distribution $F$ in $\mathbb{R}$ and any function $\psi(\cdot)$ in $\mathcal{H}$, there exists a wFCS distributed random element $X(\cdot)$ such that $\langle X, \psi\rangle \sim F$.
\end{proposition}

\subsection{Theoretical properties of PASS} \label{sec meth subsection: theory of PASS}

The first result in this subsection shows that, restricted to the wFCS distribution class, the PASS covariance function \eqref{sec meth eq: pass} has the same set of eigenfunctions as those of the regular covariance function \eqref{sec meth eq: cov fun}. Additionally, although the eigenvalues of PASS are generally different from those of the regular covariance function, they remain of the same order under an additional distributional assumption.
\begin{theorem} \label{sec meth thm: Functional Oja}
    Let $X(\cdot)$ be a random function in $\mathcal{H}$ with mean function $\mu(\cdot)$ and covariance function $\Gamma(\cdot,\cdot)$ as defined in \eqref{sec meth eq: cov fun} and $\tilde X(\cdot)$ be an independent copy of it. If $X(\cdot)$ is wFCS, then the PASS covariance function \eqref{sec meth eq: pass} admits the following decomposition
    \begin{equation}  \label{sec meth eq: Functional Oja}
	K(s,t) = \sum_{j=1}^{\infty} \lambda_j(K) \phi_j(s)\phi_j(t), ~~\text{ for any }s,t\in[0,1],
	\end{equation}
	where for any $j$ such that $\lambda_j(\Gamma)>0$, 
	\begin{align} \label{sec meth eq: pivotal eigenvalue equation}
	    \lambda_j(K) = \mathbb{E}\left[ \frac{\lambda_j(\Gamma) U_j^2 }{\sum_{i=1}^\infty \lambda_i(\Gamma) U_i^2}\right] ~\text{ with }~ U_j =  \frac{\langle X - \tilde{X},\ \phi_j \rangle}{\{2\lambda_j(\Gamma)\}^{1/2}} \text{ and } \sum_{j = 1}^\infty \lambda_j(K)=1.
	\end{align}
    If $U_j,\ j = 1,\ 2, \dotsc$ are further assumed to be exchangeable, then for any positive integers $j,\ k$, $ \lambda_j(\Gamma) \leq \lambda_k(\Gamma) $ implies that $ \lambda_j(K) \leq \lambda_{k}(K) $. In particular, if $ X - \tilde{X} $ is FE distributed, then $ U_j,\ j= 1,\ 2,\ \dotsc $ are exchangeable and follow a standard Gaussian process.
\end{theorem}
Equation \eqref{sec meth eq: Functional Oja} provides the needed theoretical support to use $\hat{K}(\cdot,\cdot)$ as a robust replacement of $\hat{\Gamma}(\cdot,\cdot)$ for eigenfunction estimation. In addition, Equation \eqref{sec meth eq: pivotal eigenvalue equation} gives a relationship between the eigenvalues of $K(\cdot,\cdot)$ and $\Gamma(\cdot,\cdot)$. However, the explicit formula linking the eigenvalues of $K(\cdot, \cdot)$ to these of $\Gamma(\cdot,\cdot)$ is complicated,  nonlinear and only known under very restrictive conditions
(cf. Proposition 3 in \citeauthor*{Durre2016a}, \citeyear{Durre2016a}
and results  in \citeauthor*{Durre2017}, \citeyear{Durre2017} for some analytical and numerical results under an additional elliptical assumption). That said, a ``conservative" signal-preservation property of eigenvalues from the PASS covariance function appears to exist, which we shall present in the following theorem. Of note, this theorem has its trace back to Proposition 2 of \cite{Durre2016a}, extending the latter to non-elliptical and also functional distributions.
\begin{theorem}[A generalization of Proposition 2 in \citeauthor{Durre2016a}, \citeyear{Durre2016}] \label{sec meth thm: eigenratio decay speed}
Assume $X(\cdot)$ to be wFCS and further that the $ U_{j},\ j = 1, 2, \dotsc $ are exchangeable. Then for any $1\leq j \leq k$, 
\begin{align}  \label{sec meth thm: eigen ratio relation}
    \frac{\lambda_j(K)}{\lambda_k(K)} \leq \frac{\lambda_j(\Gamma)}{\lambda_k(\Gamma)}.
\end{align}
\end{theorem}
Theorem \ref{sec meth thm: eigenratio decay speed} shows that $\lambda_j(K)$'s decrease at a slower rate than $\lambda_j(\Gamma)$'s do. As an immediate consequence of it, the following corollary provides guidance on rank selection in conducting FPCA; see Algorithm \ref{sec meth alg: eigenvalue estimation 2} ahead in Section \ref{subsection: Eigenvalue estimation} for its implementation.  
\begin{corollary} \label{sec meth cor: eigen ratio relation}
With the assumptions given in Theorem \ref{sec meth thm: eigenratio decay speed}, for any integer $Q \geq 1$, we have 
\begin{align*}
    \frac{\sum_{i = 1}^Q \lambda_{i}(K)}{\sum_{i = 1}^\infty \lambda_{i}(K)} \leq 
    \frac{\sum_{i = 1}^Q \lambda_i(\Gamma)}{\sum_{i = 1}^\infty \lambda_i(\Gamma)}.
\end{align*}
\end{corollary}

\section{Eigenratio estimation} \label{subsection: Eigenvalue estimation}

This section aims to recover the relative size of the top few components in the sequence  $\{\lambda_{j}(\Gamma)\}$ (a.k.a., eigenratio estimation) from Equation \eqref{sec meth eq: pivotal eigenvalue equation}, thus to obtain robust eigenvalue estimators. Even though an exact expression for $\{\lambda_{j}(\Gamma)\}$ in terms of $\{\lambda_{j}(K)\}$ in general does not exist, expressing $\lambda_j(\Gamma)$ in terms of $\lambda_j(K)$ can be regarded as solving the system of equations 
\begin{align} \label{sec them eq: sys equations}
    \lambda_j(K) = & \mathbb{E}\left[ \frac{\lambda_j(\Gamma) U_j^2}{\sum_{i=1}^\infty \lambda_i(\Gamma) U_i^2}\right],~~~\ j = 1, 2, \dotsc.
\end{align}
However, there is always one less condition than the number of unknowns. Thus, it is impossible to recover the exact value of $\lambda_j(\Gamma)$ from $\lambda_j(K)$. The intuition behind this is that $K(\cdot,\cdot)$ discards the magnitude information in favor of robustness, but eigenvalue $\lambda_j(\Gamma)$ measures the amount of variation in the direction of $\phi_j$, which requires the knowledge of how far each data point lies in that direction. As a result, $\lambda_j(K)$ does not carry sufficient information to fully estimate $\lambda_j(\Gamma)$. Nonetheless, a closer look at Equation \eqref{sec meth eq: pivotal eigenvalue equation} reveals that $\lambda_j(K)$ contains relative size information of $\lambda_j(\Gamma)$ to the other eigenvalues. In many applications of principal component analysis, the eigenvalues are mainly used to perform rank selection, and thus calculating the relative sizes of the eigenvalues $\lambda_j(\Gamma)$ suffices. 

The rest of this subsection demonstrates our approach to recovering the relative size of $\lambda_j(\Gamma)$ from Equation \eqref{sec them eq: sys equations}. Since one of these equations in \eqref{sec them eq: sys equations} is redundant, we can divide every equation by the first one (i.e., $j=1$), and noticing that obtaining the relative size among $\lambda_j(\Gamma)$ is equivalent to finding the ratios $\lambda_j(\Gamma)/\lambda_1(\Gamma),\ j = 2,\ 3, \dotsc$, without loss of generality we set $\lambda_1(\Gamma) = 1$ to get
\begin{align} \label{sec meth eq: exact sys equations}
    \lambda_j(\Gamma) = & \frac{\lambda_j(K)}{\lambda_{1}(K)} \frac{\mathbb{E} \frac{ U_1^2}{U_1^2 + \sum_{i=2}^\infty \lambda_i(\Gamma) U_i^2}}{\mathbb{E} \frac{ U_j^2}{U_1^2 + \sum_{i=2}^\infty \lambda_i(\Gamma) U_i^2}},~~~ j = 2,\ 3, \dotsc
\end{align}

In practice, the first, say, $Q$ eigenvalues will explain most of the variation. Then with $\sum_{i=2}^\infty \lambda_i(\Gamma) U_i^2 \approx \sum_{i=2}^Q \lambda_i(\Gamma) U_i^2$ and $\lambda_j(K)$ estimated by $\lambda_{j}(\hat{K})$, we have $Q$ equations with $Q$ unknowns
\begin{align} \label{sec meth eq: eigenratio equations 2}
    & \lambda_j(\Gamma) \approx \frac{\lambda_{j}(K)}{\lambda_{1}(K)} \frac{\mathbb{E} \frac{ U_1^2}{U_1^2 + \sum_{i=2}^Q \lambda_i(\Gamma) U_i^2}}{\mathbb{E} \frac{ U_j^2}{U_1^2 + \sum_{i=2}^Q \lambda_i(\Gamma) U_i^2}},~~~ j = 2, \dotsc, Q. 
\end{align}

{\color{black} The selection of $ Q $ is an important question in functional data analysis, as it is for Equations \eqref{sec meth eq: eigenratio equations 2}. For example, when the estimated eigenfunctions are used as basis functions to span the sample curves, the size of $ Q $ controls the smoothness of the spanned curves. Smooth curves resulted from a small $ Q $ are usually desirable; however, at the same time, one does not want to loss too much signal by using too small a $ Q $ \citep{Ramsay2005}. If we use $ Q $ that is not large enough, then the approximation in Equation \eqref{sec meth eq: eigenratio equations 2} will suffer; however, if we pick a $ Q $ that is too large, then unnecessary estimation errors are likely to be introduced. A common practice is to use the first $ Q $ eigenfunctions such that their corresponding eigenvalues account for certain percentage of variation, similar to the case in multivariate principal component analysis. This works well when data are not contaminated with outliers, and the sample covariance operator provides good estimates of the eigenvalues. However, when the estimated eigenvalues from the sample covariance operator are no longer reliable due to outliers, caution is required. Therefore, one cannot solely rely on the eigenvalues of the sample covariance operator when selecting the $ Q $. Definitive theory for selecting $ Q $ does not yet exist, nonetheless} a starting point for $ Q $ can be found by using Corollary \ref{sec meth cor: eigen ratio relation}. For example, if one wishes to find the first $ Q $ eigenfunctions, so that after projecting to them, at least $ 95\% $ of the variation is preserved, then a conservative starting point for the $Q$ can be the one where the cumulative percentage of variation explained (CPVE) by $\lambda_{1}(\hat{K}), \dotsc, \lambda_{Q}(\hat{K})$ is around $95\%$. Then Corollary \ref{sec meth cor: eigen ratio relation} guarantees that the first $ Q $ eigenvalues $ \lambda_{1}(\Gamma), \dotsc, \lambda_{Q}(\Gamma)$ account for at least $ 95\% $ of the variation. 

Given estimates of $\lambda_j(K)$'s and leveraging \eqref{sec meth eq: eigenratio equations 2} further, $\lambda_j(\Gamma)$'s could be estimated via a proper root finding algorithm. In the following, we elaborate on one such algorithm. Some additional notations are needed first. Define the function $g: [0,1]^{Q-1}\to \mathbb{R}^{Q-1}$ as follows:
\begin{align}
&g(x)=\Big(g_1(x),~\ldots,~g_{Q-1}(x)  \Big)^\top \label{sec meth: population fixed point funciton} \\
\text{with }~~~ & g_{i}(x) = g_i(x_1,\ldots,x_{Q-1}) := \frac{\lambda_{i+1}(K)}{\lambda_{1}(K)} \frac{\mathbb{E} \frac{ U_1^2}{U_1^2 + \sum_{l=1}^{Q-1} x_{l} U_{l+1}^2}}{\mathbb{E} \frac{ U_{i+1}^2}{U_1^2 + \sum_{l=1}^{Q-1} x_{l} U_{l+1}^2}},~~~ i = 1, \dotsc, Q-1. \nonumber
\end{align}
By \eqref{sec meth eq: eigenratio equations 2}, $ x^{*} := (\lambda_{2}(\Gamma), \dotsc, \lambda_{Q}(\Gamma))^{\top} $ is an approximated fixed point of $ g(x) $, and we thus reduce the problem to finding the fixed point of $ g(x) $. 

Even assuming further a functional elliptical distribution, the closed form for the expectations $ \mathbb{E} [U_{j}^{2} / (U_{1}^{2} + \sum x_{l}U_{l+1}^{2})] $ in $ g(x) $ has been difficult to obtain in general. An one-dimensional integral representation due to \cite{Durre2016a} however does exist:
\begin{align} \label{sec meth: integral closed form}
	\mathbb{E} \left[\frac{U_{j}^{2}}{U_{1}^{2} + \sum_{l=1}^{Q-1} x_{l} U_{l+1}^{2}} \right] = \frac{1}{2} \int_{0}^{\infty} \frac{1}{(1+x_{j} v) \prod_{l=1}^{Q-1} (1+x_{l} v)^{1/2}} dv,~~~ j = 1, \dotsc, Q-1,
\end{align}
and a multivariate eigenratio estimation algorithm based on this representation is implemented in the SSCOR package \citep[cf. Page 4 in][]{Durre2016} for elliptically distributed data.

Since the wFCS is much larger than the functional elliptical distribution, it is challenging to obtain a closed form \eqref{sec meth: integral closed form}. On the other hand, these expectations in $g(x)$ can be approximated by their corresponding sample means via a standard Monte Carlo method. This yields the following eigenratio estimation algorithm, which exploits the robustness of $\{ \phi_{j}(\hat{K}) \}$ and calculates $ U_{i} $ according to its definition given in Theorem \ref{sec meth thm: Functional Oja} without resorting to an FE distribution requirement.
\vspace{0.2cm}

\begin{algorithm}[H] \label{sec meth alg: eigenvalue estimation 2}
	\SetAlgoLined
	for a relatively large $Q$, calculate $\lambda_1(\hat{K}), \dotsc, \lambda_Q(\hat{K})$\;
	for $ l=1, \dotsc, Q $, calculate $ V_{ij, l} = s_{l}^{-1/2} \langle x_{i} - x_{j},\  \phi_{l}(\hat{K})  \rangle,\ 1 \leq i < j \leq n$, where $ s_{l} = \frac{2}{n(n-1)}  \sum_{1 \leq j < k \leq n} \langle x_{j} - x_{k},\  \phi_{l}(\hat{K})  \rangle^{2}$\;
	initialize $\Lambda^{(0)} := (\lambda_1^{(0)}, \dotsc, \lambda_Q^{(0)})^\top$\;
	\While{$\| \Lambda^{(a+1)} - \Lambda^{(a)} \| > \delta$}{
		Calculate $f_{k} \{\Lambda^{(a)} \} := \frac{1}{n(n-1)} \sum_{1 \leq i < j \leq n} \frac{V_{ij,k}^2}{V_{ij,1}^2 + \sum_{l=2}^Q \lambda^{(a)}_l V_{ij,l}^2},\ k = 1, \dotsc, Q$\;
		Update $\lambda_k^{(a+1)} = \frac{\lambda_k(\hat{K})}{\lambda_1(\hat{K})} \frac{f_{1} \{\Lambda^{(a)}\}}{f_{k} \{\Lambda^{(a)}\}},\ k = 1,\ \dotsc, Q$\;
		Calculate $\| \Lambda^{(a+1)} - \Lambda^{(a)} \|$\;
	}
	\caption{Eigenratio estimation (Beyond FE distribution)}
\end{algorithm}

Even though the $V_{ij,l}$'s in Algorithm \ref{sec meth alg: eigenvalue estimation 2} inherit the robustness from $ \{\phi_{l}(\hat{K})\} $, they might still be affected by outliers coming from $ x_{i}-x_{j} $. As a remedy, one may trim out some of the $ \langle x_{i} - x_{j},\ \hat{\phi}_{l}\rangle $'s that have the largest absolute values, and the amount to trim can be treated as a tuning parameter. Our experience in the simulation study suggests that trimming between $ 1\% $ and $ 5\% $ usually works well. A discussion on the convergence of Algorithm \ref{sec meth alg: eigenvalue estimation 2} is given in the Appendix B. Eigenratios calculated from the eigenvalues of the sample covariance operator $ \{ \lambda_{l}(\hat{\Gamma}) \} $ can be used as initial values. Though we do not have detailed information about convergence speed, Algorithm \ref{sec meth alg: eigenvalue estimation 2} converges reasonably fast with these eigenratios calculated from $ \{\lambda_{l}(\hat{\Gamma})\} $ as initial values in our simulation studies. 

\section{Extension to functional data measured with noise} \label{subsection; extension to measurement with noise}

In practice, functional observations could be recorded with measurement errors and this section extends the proposed method to handle this problem. Assuming observations on a regular grid $\{t_j=j/N, ~j=0,1,\ldots,N\}$, a widely used approach to modeling contaminated functional data is
\begin{equation} \label{sec meth eq: model with noise}
	Y(t_{j}) = X(t_{j}) + \epsilon(t_{j}),~~~ j = 1, \dotsc, N,
\end{equation}
where $ X(\cdot) $ is the true underlying process and $ \epsilon(t_{j})'s $ are the measurement errors that are mean-zero, with finite variance $\sigma^2$, independent and identically distributed (i.i.d.), and further assumed to be independent of $ X(\cdot) $. Define the PASS covariance function of $ \{Y(t_{j}),\ j = 1, \dotsc, N\} $ as 
\begin{equation} \label{sec meth: PASS with noise def}
	K^{*} (t_{j}, t_{l}) = \mathbb{E}\left[ \frac{[Y(t_{j})-\widetilde{Y}(t_{j})][Y(t_{l})-\widetilde{Y}(t_{l})]}{\| Y_{p}(\cdot)-\widetilde{Y}_{p}(\cdot) \|^{2}}\right],~~\text{ for } 1\leq j, l\leq N.
\end{equation}
Here the denominator is the norm of a piecewise constant function $ Y_{p}(\cdot) - \widetilde{Y}_{p}(\cdot)$ constructed from $\{Y(t_j)\}$ where for each $t \in [0, t_1]$ 
\[
Y_p(t):= X_{p}(t) + \epsilon_{p}(t),\quad X_p(t) := X(t_1),\quad \epsilon_p(t):=\epsilon(t_1),
\]
and for each $t \in (t_{j-1}, t_{j}]$ with $j =2 , \dotsc, N$
\[
Y_p(t):= X_{p}(t) + \epsilon_{p}(t),\quad X_p(t) := X(t_{j}),\quad \epsilon_p(t):=\epsilon(t_{j}).
\]
We further define $ \tilde{Y}_{p}(\cdot),\ \tilde{X}_{p}(\cdot),\ \tilde{\epsilon}_{p}(\cdot) $ similarly with $\{\tilde{X}(t_{j}), j=1,\ldots,N\} $ and $ \{\tilde{\epsilon}(t_{j}), j=1,\ldots,N\}$ as independent copies of $\{X(t_{j}), j=1,\ldots,N\}$ and $\{\epsilon(t_{j}), j = 1, \ldots, N\}$ respectively.
\begin{theorem} \label{sec meth thm: functional Oja with noise}
	With observations $ Y(t_{j}),\ j = 1, \dotsc, N $ on a regular grid over $[0,1]$ as given in \eqref{sec meth eq: model with noise}, if $ X(\cdot)$ is wFCS and further $\mathbb{E}\{\|X(\cdot) - \tilde{X}(\cdot)\|^{-2} \} < \infty$,
	then as $ N \rightarrow \infty $, $ K^{*} (t_{j}, t_{l}) $ admits the following decomposition
	\begin{equation*}
		K^{*}(t_{j}, t_{l}) = \sum_{i=1}^{\infty} \lambda_{i}(K^{*}) \phi_{i}(t_{j}) \phi_{i}(t_{l}) +  C 1(j=l) + o(1),
	\end{equation*}
	where $ \phi_{i}(t),\ i=1,\ 2, \dotsc $ are eigenfunctions of the covariance function $ \Gamma(s,t)$,
	\begin{equation*}
	\lambda_{i}(K^{*}) = \mathbb{E} \left[\frac{\lambda_{i}(\Gamma) U_{i}^{2}}{\| Y_{p}(\cdot) - \widetilde{Y}_{p}(\cdot)\|^{2}}\right],	
	\end{equation*}
with $ U_{i} := \lambda_{i}(\Gamma)^{-1/2}  \langle X - \widetilde{X},\ \phi_{i} \rangle$ and 
\begin{align*}
	C:=\mathbb{E}\left[ \frac{[\epsilon(t_{1})-\widetilde{\epsilon}(t_{1})]^{2}}{\| X(\cdot) - \widetilde{X}(\cdot) \|^{2} + \| \epsilon_{p}(\cdot) - \widetilde{\epsilon}_{p}(\cdot) \|^{2}}\right].
\end{align*}
If $ U_{i},\ i = 1,\ 2, \dotsc$ are further assumed to be exchangeable, then for any $ j,\ k \in \mathbb{N} $, $ \lambda_{j}(\Gamma) \leq \lambda_{k}(\Gamma) $ implies that $ \lambda_{j}(K^{*}) \leq \lambda_{k}(K^{*}) $. 
\end{theorem}
Theorem \ref{sec meth thm: functional Oja with noise} shows that with noisy sample, the PASS covariance function is similar to the expectation of the so called ``raw'' covariance function given in \cite{yao2005functional}:
\begin{align*}
	\Gamma^{*}(t_{j}, t_{l}) & := \mathbb{E}[X(t_{j}) - \hat{\mu}(t_{j})][X(t_{l}) - \hat{\mu}(t_{l})]   = \sum_{i=1}^{\infty} \lambda_{i} \phi_{i}(t_{j}) \phi_{i}(t_{l}) + \sigma^{2} 1(j=l) + o(1), 
\end{align*}
where $\hat \mu(\cdot)$ denotes the estimated mean by a local linear smoother with the form of Equation (A.2) in \cite{yao2005functional}. The similarity here is in the sense that the first parts of $ K^{*}(t_{j}, t_{l}) $ and $ \Gamma^{*}(t_{j}, t_{l}) $ share the same eigenfunctions $ \phi_{i}(\cdot) $, which are also the true eigenfunctions of $\Gamma(t_j, t_l)$, and the second part is a diagonal matrix. The only difference is that the PASS covariance function sacrifices the exact size information of the eigenvalues in favor of the robustness; however, the relative size information of the eigenvalues, which is more relevant in practice, is still preserved. More importantly, Theorem \ref{sec meth thm: functional Oja with noise} suggests that with noisy observations $ y_{i}(t_{j}) = x_{i}(t_{j}) + \epsilon_i(t_{j}),\ j = 1, \dotsc, N,\ i = 1,\ \dotsc,\ n $, existing smooth FPCA methods based on the sample covariance function $$\hat{\Gamma}_{Y}(t_{j}, t_{l}) = \frac{1}{n-1} \sum [y_{i}(t_{j}) - \bar{y}(t_{j})] [y_{i}(t_{l}) - \bar{y}(t_{l})]$$ can be directly employed with the sample PASS covariance function 
\begin{align} \label{sec meth eq: sample psscf with noise}
\hat{K}_{Y}(t_{j},t_{l}) := \frac{2}{n(n-1)} \sum_{1 \leq i < k \leq n} \frac{ \{y_i(t_{j}) - y_k(t_{j})\} \{y_i(t_{l}) - y_k(t_{l})\} }{ \| y_{p,i}(\cdot) - y_{p,k}(\cdot) \|^2 }
\end{align}
to replace the sample covariance function for achieving robustness. Some examples of the $ \hat{\Gamma}_{Y}(t_{j},t_{l}) $-based smooth FPCA methods include pre-smoothing the sample curves as suggested in \cite{Ramsay1991}, using roughness penalties as in \cite{rice1991estimating} and \cite{Silverman1996}, and smoothing the sample covariance functions as in \cite{staniswalis1998nonparametric} and \cite{yao2005functional}. All of the above methods can be extended to obtain robust smooth FPCA methods that is based on $ \hat{K}_{Y}(t_{j},t_{l}) $, and in particular Theorem \ref{sec meth thm: functional Oja with noise} indicates that when smoothing the sample PASS covariance function, the diagonal components should be removed. \cite{staniswalis1998nonparametric} and \cite{yao2005functional} made similar observations for smoothing the sample version of $ \Gamma^{*}(t_{j},t_{l}) $.

\section{Simulation} \label{sec simulation}

This section studies the finite sample performance of the eigenfunction and eigenratio estimation procedures based on the proposed PASS covariance function. Two scenarios based on whether sample curves are observed with noise are considered. For the no noise scenario, the data are observed at $i = 1, \dotsc, 101$ equally spaced points $t_i \in [0,1]$ and generated from the Karhumnen-Lo\`eve expansion \eqref{sec meth eq: KL expansion} of four eigenfunctions with nonzero corresponding eigenvalues, specified as the Fourier bases $\phi_1(t) := \sqrt{2} \sin(2\pi t),\ \phi_2(t) := \sqrt{2} \cos(2\pi t),\ \phi_3(t) := \sqrt{2} \sin(4\pi t),~ \text{ and } \phi_4(t) := \sqrt{2} \cos(4\pi t)$. The mean function is specified as $ \mu(t) := 2t(1-t).$ The random scores $\xi_j$ are generated with $\mathbb{E} (\xi_j) = 0$ and $\var(\xi_j) = \lambda_j,\ j = 1, \dotsc, 4$, where $(\lambda_1,\ \lambda_2,\ \lambda_3,\ \lambda_4)^\top = (2,\ 1,\ 1/2,\ 1/4)^\top$ is the vector of eigenvalues. 

Specifically, the scores $(\xi_1,\ \xi_2,\ \xi_3,\ \xi_{4})^\top := (\sqrt{\lambda_1}S_1,\ \sqrt{\lambda_2}S_2,\ \sqrt{\lambda_3}S_3,\ \sqrt{\lambda_4}S_{4})^\top$ are generated as follows. Hereafter for any random variable $X$, ${\rm sd}(X)$ represents the standard deviation of $X$. 1) $S_{1}, \dotsc, S_{4}$ are mutually independent and each follows the standard normal distribution; 2) $ S_j = (Z_j-\mathbb{E}Z_j)/{\rm sd}(Z_j),\ j = 1, \dotsc, 4$ and $Z_j$'s are independently generated from the Frechet distribution with location 0, scale 2, and shape 3; 3) $S_j = (Z_j-\mathbb{E}Z_j)/{\rm sd}(Z_j),\ j = 1, \dotsc, 4$ and $Z_j$'s are independently generated from the log normal distribution with location 0 and scale 2; 4) $S_j=(Z_j-\mathbb{E}Z_j)/{\rm sd}(Z_j),\ j = 1, \dotsc, 4$ and $Z_j$'s are independently generated from the Chi square distribution with degree of freedom 1; 5) Generate $(\xi_1,\ \xi_2,\ \xi_3,\ \xi_{4})$ from multivariate-T distribution with degrees of freedom $5$ and a covariance matrix that is diagonal with diagonal values $\lambda_1,\ \lambda_2,\ \lambda_3,\ \lambda_4$.

To examine the finite sample robustness of the proposed method, we also examine situations where $5\%$ of the simulated data are contaminated by outliers. Two sources of outliers are considered. One is through the mean function, where we add a shift of $ 5 $ to the mean functions of $ 5\% $ of the data (OL1).  The other type is through the score and the eigenfunction, where $3\sqrt{\lambda_1}$ is added to the first score $ \xi_{1} $ and the first eigenfunction $ \phi_1(t) $ is replaced by a simple linear function $f(t)=t$ on $ 5\% $ of the data (OL2). 

In addition to the above no noise setting, we also studied the scenario where the sample curves are observed with noise. In this case, the data generation procedure is the same as above except that at each observed point, we add a mean $0$ normal random noise with standard deviations assigned to be $0.5,\ 1,\ 2$ in three different settings. We do not add extra outliers here. In each setting described above, $1,000$ samples of sizes $100,\ 200,\ 400,\ 800$ are taken. 

Under the setting with no observation noise, we consider the following methods for comparison. For eigenfunction estimation, 1) PASS$_f$: the pairwise spatial sign  method; 2) Cov$_f$: the covariance function based method that solves for eigenfunctions of the sample covariance function given by Equation \eqref{sec meth eq: sample cov fun}; 3) MSPC: the median spherical principal component method that is proposed by \cite{Gervini2008}. For eigenratio estimation, 1) PASS$_{vm}$: 
the Monte Carlo based Algorithm \ref{sec meth alg: eigenvalue estimation 2} 
given in Section \ref{subsection: Eigenvalue estimation}; 2) PASS$_{ve}$: 
the elliptical distribution induced integral representation (Equation \ref{sec meth: integral closed form}); see also SSCOR package (cf. page 4 in \citeauthor{Durre2016}, \citeyear{Durre2016}); 3) Cov$_v$: eigenratio estimation method based on the sample covaraince function that solves for eigenvalues of the sample covariance function (Equation \ref{sec meth eq: sample cov fun}).

In the setting with observation noise, for eigenfunction estimation, the only difference is that PASS$_f$ and Cov$_f$ are applied under two smoothing schemes: 1) Smooth-CF: under Smooth-CF, we directly smooth the sample PASS covariance function and the sample covariance function with their corresponding diagonal elements removed. The smoothing is done with bivariate penalized spline and implemented by the spm function in the SemiPar package (cf. page 26 in \citeauthor{Wand2018}, \citeyear{Wand2018}); 2) Pre-smooth: we pre-smooth the data via the smooth.spline function in the stats package (\citeauthor{RCoreTeam2020}, \citeyear{RCoreTeam2020}), and then apply PASS and Cov. 

Here, PASS$_f$ under scheme Smooth-CF is justified by Theorem \ref{sec meth thm: functional Oja with noise} in Section \ref{subsection; extension to measurement with noise} and Cov$_f$ under scheme Smooth-CF is justified by \cite{yao2005functional}, e.g., on page 579 and the references therein. For the Pre-smooth scheme, applying functional data analysis techniques to discrete data after smoothing the data is well established; see for example \cite{Ramsay2005} and references therein. MSPC is developed for observations without noise and there is no known results justifying its validity under smoothing scheme Smooth-CF; nonetheless, we apply this method under scheme Pre-smooth to study its empirical performance with noisy observations. For eigenratio estimation when data are observed with noise, the three eigenratio estimation procedures mentioned in the no observation noise setting, namely PASS$_{vm}$, PASS$_{ve}$, and Cov$_v$ are compared under the two smoothing schemes: Smooth-CF and Pre-smooth.

The performance of the eigenfunction estimation is measured by the sample version of the mean integrated square errors (MSE) $ \mathbb{E}\| \hat{\phi}_1(\cdot) - \phi_1(\cdot) \|^2$ and the sample version of the bias 
$\| \mathbb{E} \hat{\phi}_1(\cdot) - \phi_1(\cdot) \|$. The performance of the eigenratio estimation is measured via the mean squared error (MSE) of the estimated percentage of variance explained (PVE) by the first estimated eigenvalue.

\begin{table}
\caption{Estimated MSE $\times 10^{-2}$ of the first eigenfunction and its estimated bias $\times 10^{-2}$ when sample curves are observed without noise. Sample size is 200. The three methods are pairwise spacial sign (PASS$_{f}$), covariance function based (Cov$_{f}$), and median spherical principal component (MSPC). Table is divided into three parts corresponding to outlier types. Top portion (OL0) shows results where no outliers are added followed by results where 5\% of the data have additional outliers on their mean function (OL1), and the bottom portion shows results where 5\% of the data have additional outliers on their first scores and eigenfunctions (OL2).} 
\label{simulation eigenfun on all socres and eigenfunctions}
{\begin{tabular}{rcccccc}
  \hline
 & \multirow{2}{7em}{Methods$\backslash$Distr }   & Normal & Mult-T & Frechet & Log-normal & Chi square \\ 
  &  & MSE\ \ Bias & MSE\ \ Bias & MSE\ \ Bias & MSE\ \ Bias & MSE\ \ Bias \\
  \hline
   & PASS$_{f}$ & 1.5\ \ \ 0.0 & 1.9\ \ \ 0.0 &  2.8\ \ \ 0.0 &  3.4\ \ \ 0.0 &  4.3\ \ \ 0.0 \\ 
   OL0 & Cov$_{f}$ & 1.4\ \ \ 0.0 & 3.8\ \ \ 0.0 & 35.1\ \ \ 3.0 & 25.7\ \ \ 1.6 &  7.8\ \ \ 0.1 \\ 
    & MSPC & 2.2\ \ \ 0.0 & 2.4\ \ \ 0.0 &  4.9\ \ \ 1.1 &  7.7\ \ \ 3.4 & 10.8\ \ \ 6.3 \\ 
    \hline
   & PASS$_{f}$ & 1.6\ \ \ 0.0 & 1.9\ \ \ 0.0 &  2.9\ \ \  0.0 &  3.4\ \ \ 0.0 &  4.6\ \ \ 0.0 \\ 
    OL1 & Cov$_{f}$ & 3.1\ \ \ 0.0 & 6.4\ \ \ 0.1 & 68.0\ \ \ 11.7 & 46.5\ \ \ 5.4 & 12.1\ \ \ 0.3 \\ 
    & MSPC & 2.3\ \ \ 0.0 & 2.5\ \ \ 0.0 &  5.3\ \ \  1.2 &  8.1\ \ \ 3.5 & 11.1\ \ \ 6.4 \\ 
    \hline
   & PASS$_{f}$ & 2.7\ \ \ 0.9 & 3.3\ \ \ 1.2 &  5.6\ \ \ 2.5 &  6.7\ \ \ 2.3 &  8.7\ \ \ 1.5 \\ 
    OL2 & Cov$_{f}$ & 3.7\ \ \ 2.0 & 6.6\ \ \ 2.1 & 39.7\ \ \ 7.1 & 31.2\ \ \ 4.8 & 11.4\ \ \ 2.3 \\ 
    & MSPC & 3.4\ \ \ 0.7 & 3.5\ \ \ 0.9 &  6.6\ \ \ 1.5 & 11.0\ \ \ 3.8 & 15.7\ \ \ 8.1 \\ 
   \hline
\end{tabular}}
\end{table}

The relative performance among the compared methods are similar across the sample sizes considered; therefore, only the results with sample size 200 are presented here. Table \ref{simulation eigenfun on all socres and eigenfunctions} presents the eigenfunction estimation results for the no noise setting. The (OL0) group is where no outliers are added to the data; (OL1) and (OL2) are cases with added outliers defined above. Among the three methods compared, PASS$_f$ has the smallest estimated  MSE in all situations except under normal distribution with no outliers added, where its estimated  MSE is very close to that of Cov$_f$. In terms of bias, PASS$_f$ is similar to MSPC in most situations, however, there are some cases, MSPC's bias is much larger than that of PASS$_f$. Under the heavy-tailed distributions, PASS$_f$ and MSPC outperform Cov$_f$ by a large margin. It is worth pointing out that MSPC's non-robustness against skewed data is exemplified under the Chi square distribution. In contrast, the robustness due to the pairwise construction of PASS$_f$ in such situation is clear. 

\begin{figure}
    \centering
    \includegraphics[scale=0.54]{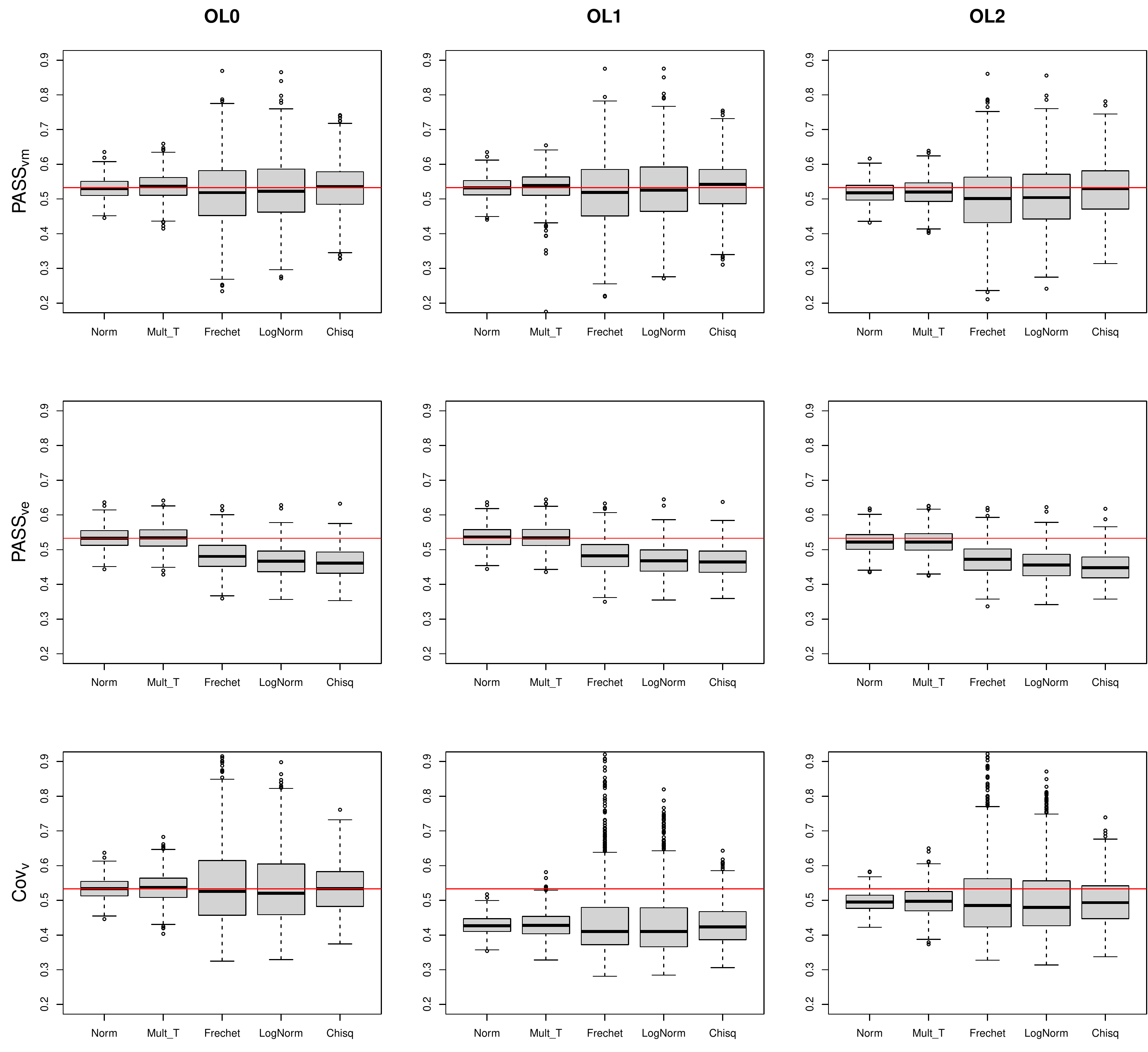}
    \caption{Boxplots of the estimated percentage of variance explained by the first eigenratio when sample curves are observed without noise. The red horizontal line indicates the true value. Sample size is 200. The first two rows are the PASS based methods, where PASS$_{vm}$ uses Algorithm \ref{sec meth alg: eigenvalue estimation 2} and PASS$_{ve}$ uses the elliptical distribution induced integral representation based algorithm in \cite{Durre2016a}. The third row is the covariance function based method (Cov$_{v}$). The first column shows results where no outliers are added (OL0) followed by the second column that shows results where 5\% outliers are added on the mean function (OL1), and the third column shows results where 5\% outliers are added on the first score and first eigenfunction (OL2).}
    \label{simulation fig: eigenvalue}
\end{figure}

Figure \ref{simulation fig: eigenvalue} shows the eigenratio estimation results when data are observed without noise. Under normal and Multivariate-T, all three methods perform similarly when no outliers are added (OL0). When outliers are present in the data, both PASS$_{vm}$ and PASS$_{ve}$ outperform Cov$_v$. Under the other three distribution settings, PASS$_{vm}$ performs similarly to Cov$_v$ when no outliers are added and outperforms Cov$_v$ when outliers are added. As mentioned in Section \ref{subsection: Eigenvalue estimation}, PASS$_{ve}$ is tailored for elliptical pairwise differences, and this is reflected by the simulation results here. For example, under normal and multivariate-T, it performs well; however, PASS$_{ve}$'s reliance on elliptical pairwise differences is evident under Frechet, Log-normal, and Chi square distributions where it way underestimates the PVE by the true first eigenvalue. The message from the eigenvalue estimation results when data are observed with noise is similar to the one when data are observed without noise; thus with noise eigenvalue estimation results are given in the Appendix C.

\begin{table}
\caption{MSE $\times 10^{-2}$ of the estimated first eigenfunction and its bias $\times 10^{-2}$ when sample curves are observed with mean $0$ standard deviation $\sigma = 0.5, 1, 2$ Normal random noise. Sample size is 200. The three methods are
pairwise spatial sign (PASS$_{f}$), covariance function based (Cov$_{f}$), and median spherical principal component (MSPC). PASS$_{f}$ and Cov$_{f}$ are applied under to smoothing schemes: 1) pre-smooth data (Pre-smooth) and 2) directly smooth the corresponding covariance functions with diagonal elements removed (Smooth-CF). MSPC is only applied under smoothing scheme 1) (Pre-smooth). The table is divided into three portions corresponding to $\sigma = 0.5,\ 1,\  2$. No additional outliers are added.} \label{simulation eigenfun with noise}
\begin{tabular}{rrccccc}
  \hline
 & \multirow{2}{6em}{Methods$\backslash$Distr} & Normal & Mult-T & Frechet & Log-normal & Chi square \\
  & & MSE\ \  Bias & MSE\ \  Bias & MSE\ \  Bias & MSE\ \  Bias & MSE\ \  Bias \\ 
  \hline
   & {\footnotesize PASS$_{f}$}:~~~~~~~~~~\\ 
   & {\footnotesize Smooth-CF} & 1.6\ \ \ 0.1 & 1.9\ \ \ 0.0 &  3.1\ \ \ 0.0 &  4.7\ \ \ 0.3 &  4.5\ \ \ 0.0 \\ 
    & {\footnotesize Pre-smooth} & 1.6\ \ \ 0.1 & 1.8\ \ \ 0.0 &  2.8\ \ \ 0.0 &  4.4\ \ \ 0.3 &  4.4\ \ \ 0.0 \\
    $\sigma=0.5$ & {\footnotesize Cov$_{f}$}:~~~~~~~~~~~~\\ 
     &{\footnotesize Smooth-CF} & 1.5\ \ \ 0.1 & 3.8\ \ \ 0.0 & 35.1\ \ \ 3.0 & 26.5\ \ \ 1.7 &  7.8\ \ \ 0.1 \\ 
     &{\footnotesize Pre-smooth} & 1.5\ \ \ 0.1 & 3.8\ \ \ 0.0 & 35.3\ \ \ 3.0 & 26.8\ \ \ 1.7 &  8.0\ \ \ 0.1 \\ 
      & {\footnotesize MSPC}:~~~~~~~~~~\\
      &{\footnotesize Pre-smooth} & 2.3\ \ \ 0.1 & 2.4\ \ \ 0.0 &  4.9\ \ \ 1.2 &  9.1\ \ \ 4.6 & 10.4\ \ \ 5.9 \\ 
    \hline
  & {\footnotesize PASS$_{f}$}:~~~~~~~~~~\\
  &{\footnotesize Smooth-CF} & 1.8\ \ \ 0.2 & 1.9\ \ \ 0.0 &  3.8\ \ \ 0.0 &  6.3\ \ \ 0.8 &  4.6\ \ \ 0.0 \\ 
    & {\footnotesize Pre-smooth} & 1.9\ \ \ 0.3 & 1.9\ \ \ 0.0 &  3.2\ \ \ 0.1 &  6.2\ \ \ 1.4 &  4.7\ \ \ 0.1 \\  
   $\sigma=1$ &{\footnotesize Cov$_{f}$}:~~~~~~~~~~~~\\
    & {\footnotesize Smooth-CF} & 1.7\ \ \ 0.2 & 3.8\ \ \ 0.0 & 35.1\ \ \ 3.0 & 27.2\ \ \ 1.8 &  7.9\ \ \ 0.1 \\ 
    & {\footnotesize Pre-smooth} & 1.8\ \ \ 0.3 & 3.9\ \ \ 0.0 & 35.5\ \ \ 3.1 & 27.8\ \ \ 2.0 &  8.2\ \ \ 0.1 \\
    & {\footnotesize MSPC}:~~~~~~~~~~\\
    &{\footnotesize Pre-smooth} & 2.7\ \ \ 0.3 & 2.5\ \ \ 0.1 &  5.3\ \ \ 1.4 & 11.8\ \ \ 7.0 & 10.1\ \ \ 5.4 \\
    \hline
   & {\footnotesize PASS$_{f}$}:~~~~~~~~~~\\
   & {\footnotesize Smooth-CF} & 2.7\ \ \ 1.0 & 2.2\ \ \ 0.0 &  6.8\ \ \ 0.1 & 10.3\ \ \  2.3 &  5.4\ \ \ 0.1 \\ 
    & {\footnotesize Pre-smooth} & 3.2\ \ \ 1.4 & 2.2\ \ \ 0.2 &  5.0\ \ \ 0.9 & 12.7\ \ \  6.7 &  6.0\ \ \ 0.4 \\  
  $\sigma=2$&{\footnotesize Cov$_{f}$}:~~~~~~~~~~~~\\
  & {\footnotesize Smooth-CF} & 2.6\ \ \ 0.9 & 4.0\ \ \ 0.0 & 34.5\ \ \ 2.9 & 28.9\ \ \  2.6 &  8.1\ \ \ 0.1 \\ 
    & {\footnotesize Pre-smooth} & 2.9\ \ \ 1.3 & 4.1\ \ \ 0.1 & 35.5\ \ \ 3.2 & 30.1\ \ \  3.0 &  8.9\ \ \ 0.3 \\
  & {\footnotesize MSPC}:~~~~~~~~~~\\
  & {\footnotesize Pre-smooth} & 4.3\ \ \ 1.6 & 3.2\ \ \ 0.7 &  9.0\ \ \ 3.7 & 22.0\ \ \ 16.5 & 11.0\ \ \ 5.4 \\ 
   \hline
\end{tabular}
\end{table}

The results for eigenfunction estimation when data are observed with noise are presented in Table \ref{simulation eigenfun with noise}. There is no clear difference between the two smoothing schemes for PASS$_f$ in terms of MSE, although PASS$_f$ does seem to hold a slight edge under Smooth-CF in terms of bias, but not by a large margin. Cov$_f$ under the Smooth-CF seems to perform better than it is under Pre-smooth, but not by very much. PASS$_f$ under both smoothing schemes perform better than the other two methods in most situations except under normal distribution, where it is close to Cov$_f$. In particular, PASS$_f$ outperforms Cov$_f$ by a large margin under Frechet and Log-normal distributions regardless the smoothing schemes.

\section{Applications}  \label{sec application}

In this section, we illustrate the robustness of the proposed PASS FPCA method via applying it to an accelerometry data in physical activity epidemiological studies. 

Accelerometers have been widely used to measure and monitor physical activity objectively in large-scale epidemiological studies in the past decade. One of these studies is an ancillary study of the Women's Health Initiative (WHI), called the Objective Physical Activity (PA) and Cardiovascular Health (OPACH) Study \citep{LaCroix2017}. The WHI consists of an observational study involving 93,676 post-menopausal women and three clinical trials involving 68,132 post-menopausal women. The clinical outcomes include cardiovascular disease, cancer, and many other health outcomes. In 2012--2014, the WHI Long Life Study was conducted to collect new data for the purpose of supporting research into factors associated with healthy aging and changing levels of intermediate markers of cardiovascular risk. As an ancillary study to the WHI Long Life Study, the OPACH collects objective and subjective physical activity measurements on more than 6,500 women, via accelerometers and self-reported questionnaires respectively. The primary aim of OPACH is to determine the associations of objectively measured PA by accelerometry with risks of incident CVD and total mortality. 

Each accelerometer records 3-dimensional accelerations of a woman 30 times at each second. Data were downloaded both in the format of raw data and counts per 15-seconds. The latter can be summarized into a vector called vector magnitude (VM) with 10,080 observations, where each observation is an activity count that measures the total physical activity in one minute for seven days.

In our analysis, the data are reduced to a subset of 6,389 women with at least one valid day, which is defined as with full 24-hour observations. Then the first valid day's VM for each woman is included in the final analysis. The data is then aggregated into one minute resolution, which results an $1440 \times 1$ vector for each woman. Therefore, our final data set contains 6,389 VM's of length 1,440, which are regarded as sample curves.

\begin{figure}
    \centering
	\includegraphics[width=15cm]{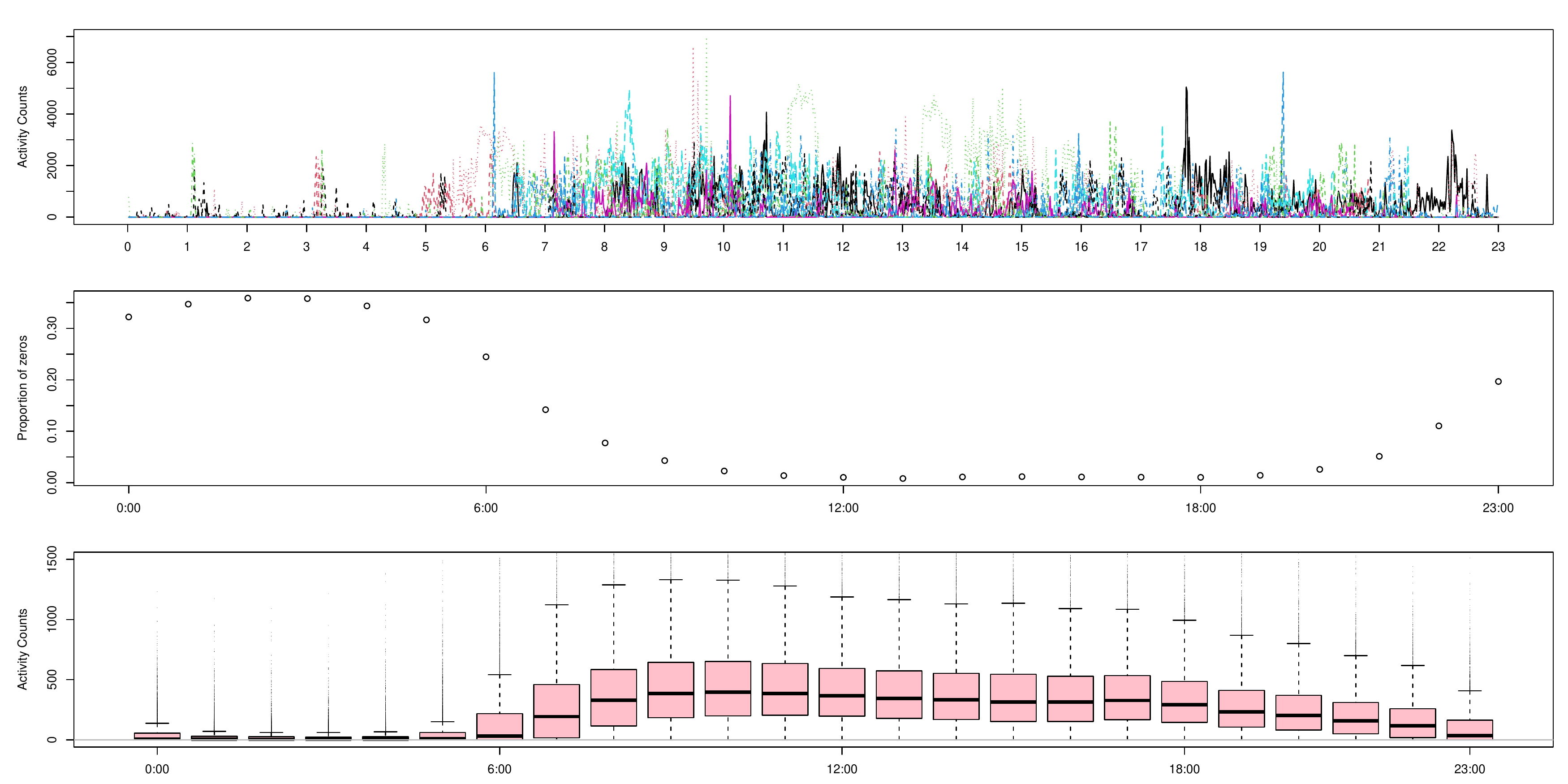}
	\caption{Plots of activity counts data. The top plot shows ten randomly selected observations. The plot in the middle shows the proportion of 0 counts at each hour. The bottom are boxplots of the activity counts data at different hours.} \label{sec app: acc data boxplots}
\end{figure}

The top plot in Figure \ref{sec app: acc data boxplots} displays an example of 10 randomly selected VM curves out of the 6389. The middle plot shows the proportion of 0 counts at each hour. The bottom of Figure \ref{sec app: acc data boxplots} contains boxplots of activity counts for these 6389 women at different hours within a day. It is evident that activity counts data are highly skewed, so that the PASS FPCA, which does not require stringent symmetry conditions, coupled with a smoothing procedure is more suitable for analysis. As indicated by the simulation results in Table \ref{simulation eigenfun with noise}, smoothing schemes Smooth-CF and Pre-smooth are very close for both PASS based FPCA and covariance based FPCA with dense data; thus, in the following we use Pre-smooth due to its relative computational convenience. 

\begin{figure}
	\begin{center}
		\begin{tabular}{c}
			\includegraphics[width=16cm]{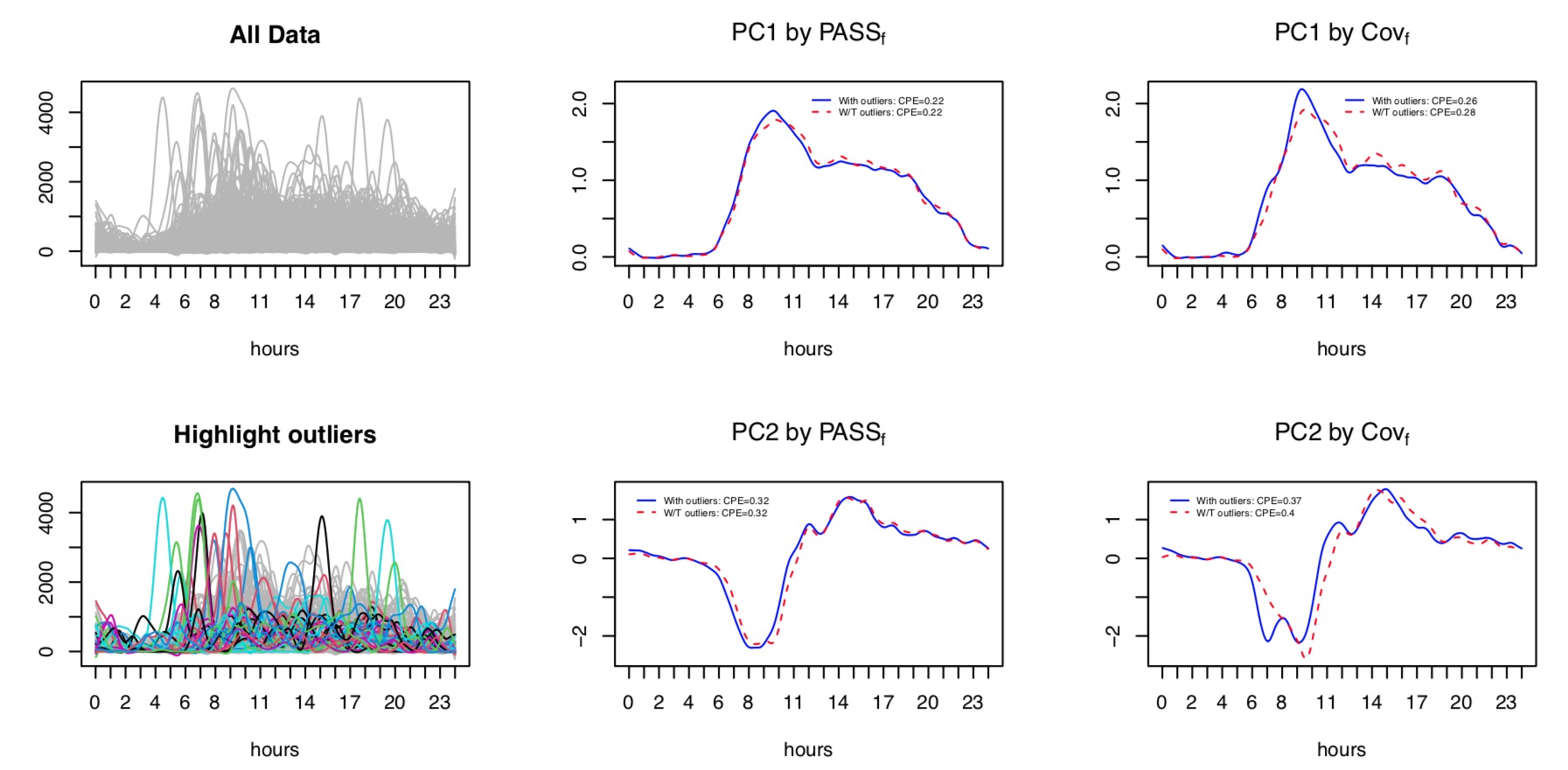}
		\end{tabular}
	\end{center}
	\caption{Accelerometery data: Estimated eigenfunctions and cumulative percentage of variation (CPE) by PASS$_f$ and Cov$_f$ with outliers included v.s. outliers removed.} 
	\label{sec app: acc results}
	\vspace{-0.1in}
\end{figure}

Due to the large number of sample curves, it is difficult to visually identify outliers. For illustrative purposes, we focus on a sample of 500 from the 6,389 VM curves, especially to demonstrate the influence of potential outliers on eigenfunction estimations. Figure \ref{sec app: acc results} displays the first two estimated eigenfunctions by PASS$_f$ and Cov$_f$. The top left figure shows the general shape of the spline smoothed 500 data, and the bottom left figure highlights 34 outliers that are identified by directional quantile that is discussed by \citet{Myllymaki2015} and implemented via the fdaoutlier package \citep*[cf page 8 in][]{Ojo2020}. By comparing the eigenfunctions estimated with and without outliers, PASS$_f$ can be seen to be very robust. However, the estimated eigenfunctions by Cov$_f$ are highly sensitive to the presence of outliers. The sensitivity is especially prominent in the morning hours, for instance, the first and second eigenfunctions estimated by Cov$_f$ have a jump around 9am and 8am respectively. However, when outliers are removed, the estimated eigenfunctions by Cov$_f$ are much closer to that estimated by PASS$_f$. In the plots, we also include the cumulative percentage of variance explained (CPE) by the principal components. It can be seen that the CPE's estimated by Algorithm \ref{sec meth alg: eigenvalue estimation 2} that is based on PASS$_{vm}$ are more robust than those estimated by Cov$_v$. The estimated first eigenfunction shows that during the early hours from midnight to 6AM, there is not much variation in activity counts. The variation climbs up during the day and peaks at around 10AM, then it starts to go down in the afternoon until midnight. Similar observations to Figure \ref{sec app: acc results} can be made when outliers are identified by other criteria such as L$^2$ norm of the data. These observations can also be made when PASS FPCA and Cov FPCA are applied to the whole 6389 curves. We relegate those results to the Appendix C. 

To further demonstrate the robustness of the PASS FPCA, we used the standardized scores, which are calculated by using the eigenfunctions estimated by PASS$_f$ and Cov$_f$ respectively, as regressors adjusted by age, BMI and ethnicity to fit a linear model with glucose level as response. 12 standardized cores are used in the model, and ethnicity includes 2550 White Americans, 1492 African Americans, and 828 Hispanic or Latino Americans.

\begin{table}
\caption{Regression results for glucose. Scores are estimated by two methods: PASS$_f$ and Cov$_f$. The linear model is fitted with the whole data and with the outliers removed respectively. Scores 8 - 12 are non-significant across PASS$_f$ and Cov$_f$ with or without outliers, thus omitted in the table.} 
\label{regression table}
{\begin{tabular}{lllll}
    \hline
    & \multicolumn{2}{c}{Scores by PASS$_f$} & \multicolumn{2}{c}{Scores by Cov$_f$} \\
  	&	All data	&	No outliers	&	All data	&	No outliers	\\
	&	Coef (P value)	&	Coef (P value)	&	Coef (P value)	&	Coef (P value)	\\
	\hline
Intercept	&	87.7 (0)	&	90.65 (0)	&	87.63 (0)	&	90.65 (0)	\\
score 1	&	2.33 (0)	&	2.36 (0)	&	2.44 (0)	&	2.44 (0)	\\
score 2	&	-0.74 (0.05)	&	-0.53 (0.18)	&	-0.75 (0.05)	&	-0.48 (0.22)	\\
score 3	&	0.19 (0.62)	&	0.08 (0.84)	&	0.38 (0.32)	&	0.09 (0.82)	\\
score 4	&	0.65 (0.09)	&	0.38 (0.34)	&	0.84 (0.03)	&	0.54 (0.17)	\\
score 5	&	-0.23 (0.56)	&	-0.04 (0.92)	&	-0.26 (0.51)	&	-0.05 (0.9)	\\
score 6	&	1.07 (0.01)	&	1.24 (0)	&	0.51 (0.18)	&	1.15 (0)	\\
score 7	&	-0.13 (0.74)	&	-0.02 (0.97)	&	-0.64 (0.09)	&	-0.21 (0.59)	\\
age	&	-0.02 (0.76)	&	-0.05 (0.49)	&	-0.02 (0.77)	&	-0.05 (0.49)	\\
bmiclls	&	3.97 (0)	&	3.82 (0)	&	3.98 (0)	&	3.82 (0)	\\
African Americans	&	-0.31 (0.76)	&	-0.38 (0.72)	&	-0.35 (0.73)	&	-0.38 (0.71)	\\
Hispanic Americans	&	1.89 (0.11)	&	1.92 (0.12)	&	1.87 (0.11)	&	1.92 (0.12)	\\
\hline
\end{tabular}}
\end{table}

Table \ref{regression table} presents the regression results. The analysis was done with all 6389 curves and compared to analysis with some outliers removed. The R package fdaoutlier \citep[][]{Ojo2020} was used to identify 283 outliers by the directional quantile method. Scores 8-12 are all non-significant for both PASS$_f$ and Cov$_f$ with or without outliers, thus they are omitted from the table. With scores estimated by PASS$_f$, the p values presented in the parenthesis give consistent conclusions about variable significance between model fitted with and without outliers. However, with scores estimated by Cov$_f$, the p values resulted by using all data and data with outliers removed give conflicting results. For example, score 4 is significant with all data but non-significant with outliers removed, and score 6 is non-significant with all data but significant with outliers removed. 

\section{Conclusion}

This paper introduced a robust FPCA method based on the functional pairwise spatial sign (PASS) covariance function. The proposed method preserves the eigenfunctions of the covariance function under a newly introduced class of distributions called the wFCS. The class of wFCSs allows for asymmetry and is strictly larger than the already well-used functional elliptical distribution class in the robust FPCA literature. Eigenratio estimation was also discussed in this paper. This paper further discussed extension of the PASS method to functional data measured with noise. For this, we showed that existing smoothing techniques developed for tradition covariance function based FPCA can be applied directly to our PASS method. The robustness of PASS is demonstrated through an intensive simulation study. An application to the accelerometry data further demonstrated the robustness of PASS.

\bibliographystyle{apalike} 
\bibliography{references.bib}

\begin{thebibliography}{}

\bibitem[Bali and Boente, 2009]{Bali2009}
Bali, J.~L. and Boente, G. (2009).
\newblock {Principal points and elliptical distributions from the multivariate
  setting to the functional case}.
\newblock {\em Statistics and Probability Letters}, 79(17):1858--1865.

\bibitem[Bali et~al., 2011]{Bali2011}
Bali, J.~L., Boente, G., Tyler, D.~E., and Wang, J.~L. (2011).
\newblock {Robust functional principal components: A projection-pursuit
  approach}.
\newblock {\em Annals of Statistics}, 39(6):2852--2882.

\bibitem[Berinde and Takens, 2007]{Berinde2007}
Berinde, V. and Takens, F. (2007).
\newblock {\em {Iterative approximation of fixed points}}.
\newblock Springer, Berinde2007.

\bibitem[Boente and Salibian-Barrera, 2015]{boente2015s}
Boente, G. and Salibian-Barrera, M. (2015).
\newblock {S-estimators for functional principal component analysis}.
\newblock {\em Journal of the American Statistical Association},
  110(511):1100--1111.

\bibitem[Boente et~al., 2014]{Boente2014}
Boente, G., {Salibi{\'{a}}n Barrera}, M., and Tyler, D.~E. (2014).
\newblock {A characterization of elliptical distributions and some optimality
  properties of principal components for functional data}.
\newblock {\em Journal of Multivariate Analysis}, 131:254--264.

\bibitem[Bosq, 2012]{Bosq2012}
Bosq, D. (2012).
\newblock {\em {Linear processes in function spaces: theory and applications
  (Vol. 149)}}.
\newblock Springer Science {\&} Business Media.

\bibitem[Cardot, 2000]{cardot2000nonparametric}
Cardot, H. (2000).
\newblock {Nonparametric estimation of smoothed principal components analysis
  of sampled noisy functions}.
\newblock {\em Journal of Nonparametric Statistics}, 12(4):503--538.

\bibitem[Castro et~al., 1986]{castro1986principal}
Castro, P.~E., Lawton, W.~H., and Sylvestre, E.~A. (1986).
\newblock {Principal modes of variation for processes with continuous sample
  curves}.
\newblock {\em Technometrics}, 28(4):329--337.

\bibitem[Cheney, 2013]{Cheney2013}
Cheney, W. (2013).
\newblock {\em {Analysis for applied mathematics}}, volume 208.
\newblock Springer Science {\&} Business Media.

\bibitem[Dauxois et~al., 1982]{dauxois1982asymptotic}
Dauxois, J., Pousse, A., and Romain, Y. (1982).
\newblock {Asymptotic theory for the principal component analysis of a vector
  random function: some applications to statistical inference}.
\newblock {\em Journal of multivariate analysis}, 12(1):136--154.

\bibitem[Deutsch, 2012]{Deutsch2012}
Deutsch, F.~R. (2012).
\newblock {\em {Best approximation in inner product spaces}}.
\newblock Springer Science {\&} Business Media.

\bibitem[D{\"{u}}rre et~al., 2017]{Durre2017}
D{\"{u}}rre, A., Fried, R., and Vogel, D. (2017).
\newblock {The Spatial Sign Covariance Matrix and Its Application for Robust
  Correlation Estimation}.
\newblock {\em Austrian Journal of Statistics}, 46:13--22.

\bibitem[D{\"{u}}rre et~al., 2016]{Durre2016a}
D{\"{u}}rre, A., Tyler, D.~E., and Vogel, D. (2016).
\newblock {On the eigenvalues of the spatial sign covariance matrix in more
  than two dimensions}.
\newblock {\em Statistics and Probability Letters}, 111:80--85.

\bibitem[D{\"{u}}rre and Vogel, 2016]{Durre2016}
D{\"{u}}rre, A. and Vogel, D. (2016).
\newblock {sscor: Robust Correlation Estimation and Testing Based on Spatial
  Signs}.
\newblock {\em R package version 0.2}.

\bibitem[Fang et~al., 1989]{Fang1989}
Fang, K.~T., Kotz, S., and Ng, K.~W. (1989).
\newblock {\em {Symmetric multivariate and related distributions}}.
\newblock CRC Press.

\bibitem[Gervini, 2008]{Gervini2008}
Gervini, D. (2008).
\newblock {Robust functional estimation using the median and spherical
  principal components}.
\newblock {\em Biometrika}, 95(3):587--600.

\bibitem[Hall and Hosseini-Nasab, 2006]{hall2006properties}
Hall, P. and Hosseini-Nasab, M. (2006).
\newblock {On properties of functional principal components analysis}.
\newblock {\em Journal of the Royal Statistical Society: Series B (Statistical
  Methodology)}, 68(1):109--126.

\bibitem[Han and Liu, 2017]{han2017eca}
Han, F. and Liu, H. (2017).
\newblock {ECA: High-Dimensional Elliptical Component Analysis in Non-Gaussian
  Distributions}.
\newblock {\em Journal of the American Statistical Association}, pages 1--17.

\bibitem[Kraus and Panaretos, 2012]{kraus2012dispersion}
Kraus, D. and Panaretos, V.~M. (2012).
\newblock {Dispersion operators and resistant second-order functional data
  analysis}.
\newblock {\em Biometrika}, 99(4):813--832.

\bibitem[LaCroix et~al., 2017]{LaCroix2017}
LaCroix, A.~Z., Rillamas-Sun, E., Buchner, D., Evenson, K.~R., Di, C., Lee,
  I.~M., and et~al. (2017).
\newblock {The Objective Physical Activity and Cardiovascular Disease Health in
  Older Women (OPACH) Study}.
\newblock {\em BMC Public Health}, 17(1).

\bibitem[Lambers and Sumner, 2018]{Lambers2018}
Lambers, J.~V. and Sumner, A.~C. (2018).
\newblock {\em {Explorations in Numerical Analysis}}.
\newblock WORLD SCIENTIFIC.

\bibitem[Locantore et~al., 1999]{Locantore1999}
Locantore, N., Marron, J., Simpson, D., Tripoli, N., Zhang, J., and K., C.
  (1999).
\newblock {Robust principal component analysis for functional data}.
\newblock {\em Test}, 8(1):1--73.

\bibitem[Marden, 1999]{Marden1999}
Marden, J.~I. (1999).
\newblock {Some robust estimates of principal components}.
\newblock {\em Statistics and Probability Letters}, 43(4):349--359.

\bibitem[Myllym{\"{a}}ki et~al., 2015]{Myllymaki2015}
Myllym{\"{a}}ki, M., Grabarnik, P., Seijo, H., and Stoyan, D. (2015).
\newblock {Deviation test construction and power comparison for marked spatial
  point patterns}.
\newblock {\em Spatial Statistics}, 11:19--34.

\bibitem[Ojo et~al., 2020]{Ojo2020}
Ojo, O.~T., Lillo, R.~E., and {Fernandez Anta}, A. (2020).
\newblock {fdaoutlier: Outlier Detection Tools for Functional Data Analysis}.
\newblock {\em R package version 0.1.1}.

\bibitem[{R Core Team}, 2020]{RCoreTeam2020}
{R Core Team} (2020).
\newblock {R: A Language and Environment for Statistical Computing}.
\newblock {\em R Foundation for Statistical Computing}.

\bibitem[Ramsay, 2006]{ramsay2006functional}
Ramsay, J.~O. (2006).
\newblock {\em {Functional data analysis}}.
\newblock Wiley Online Library.

\bibitem[Ramsay and Dalzell, 1991]{Ramsay1991}
Ramsay, J.~O. and Dalzell, C.~J. (1991).
\newblock {Some Tools for Functional Data Analysis}.
\newblock {\em Journal of the Royal Statistical Society: Series B
  (Methodological)}, 53(3):539--561.

\bibitem[Ramsay and Silverman, 2005]{Ramsay2005}
Ramsay, J.~O. and Silverman, B.~W. (2005).
\newblock {\em {Functional Data Analysis}}.
\newblock Springer Series in Statistics. Springer New York, New York, NY.

\bibitem[Rice and Silverman, 1991]{rice1991estimating}
Rice, J.~A. and Silverman, B.~W. (1991).
\newblock {Estimating the mean and covariance structure nonparametrically when
  the data are curves}.
\newblock {\em Journal of the Royal Statistical Society. Series B
  (Methodological)}, pages 233--243.

\bibitem[Silverman, 1996]{Silverman1996}
Silverman, B.~W. (1996).
\newblock {Smoothed functional principal components analysis by choice of
  norm}.
\newblock {\em Annals of Statistics}, 24(1):1--24.

\bibitem[Staniswalis and Lee, 1998]{staniswalis1998nonparametric}
Staniswalis, J.~G. and Lee, J.~J. (1998).
\newblock {Nonparametric regression analysis of longitudinal data}.
\newblock {\em Journal of the American Statistical Association},
  93(444):1403--1418.

\bibitem[Taskinen et~al., 2012]{Taskinen2012}
Taskinen, S., Koch, I., and Oja, H. (2012).
\newblock {Robustifying principal component analysis with spatial sign
  vectors}.
\newblock {\em Statistics and Probability Letters}, 82(4):765--774.

\bibitem[Wade, 2010]{Wade2010}
Wade, W.~R. (2010).
\newblock {\em {An introduction to analysis}}.
\newblock Pearson Education Inc, 4th edition.

\bibitem[Wand, 2018]{Wand2018}
Wand, M. (2018).
\newblock {SemiPar: Semiparametic Regression}.
\newblock {\em R package version 1.0-4.2}.

\bibitem[Wang et~al., 2015]{Wang2015}
Wang, J.-L., Chiou, J.-M., and Mueller, H.-G. (2015).
\newblock {Review of Functional Data Analysis}.
\newblock {\em arXiv preprint arXiv:1507.05135}.

\bibitem[Yao et~al., 2005]{yao2005functional}
Yao, F., M{\"{u}}ller, H.-G., and Wang, J.-L. (2005).
\newblock {Functional data analysis for sparse longitudinal data}.
\newblock {\em Journal of the American Statistical Association},
  100(470):577--590.

\end{thebibliography}

\renewcommand{\thesection}{A\arabic{section}}
\setcounter{section}{0}

\section{Appendix A} \label{sec proof}
This section contains proofs of the results given in the main paper. Propositions, theorems, and definitions referenced in this section are all given in the main paper with their corresponding reference numbers. We first include the multivariate elliptical distribution given by Definition 2.2 in \citet{Fang1989} and the distribution model given in \citet{Marden1999} which is larger than the multivariate elliptical distribution.

\begin{definition}[Multivariate elliptical distribution, \citeauthor{Fang1989}, \citeyear{Fang1989}]
A random vector $ X \in \mathbb{R}^{d} $ is said to follow an \emph{elliptical distribution} with parameters $ \mu \in \mathbb{R}^{d}$ and $ \Sigma \in \mathbb{R}^{d \times d} $ if $X \overset{D}{=} \mu + A^\top Y$, where ``$\overset{D}{=}$" represents ``equality in distribution", $Y \in \mathbb{R}^{k}$ is spherically distributed with characteristic function of the form $ \varphi(t^{\top}t) $ for a well-defined function $ \varphi(\cdot)$ and $ A \in \mathbb{R}^{k \times d}$, $ A^{\top}A = \Sigma $ with rank$(\Sigma)= k$. In this case, we denote $ X \sim \epsilon_{d}(\mu, \Sigma, \varphi) $ with $ \Sigma $ termed the \emph{scatter matrix} and $ \varphi(\cdot) $ the \emph{characteristic generator} of $X$. 
\end{definition}

\begin{definition} \label{sec meth: MM3}
	A random vector $ X \in \mathbb{R}^{d} $ is said to follow the model in \cite{Marden1999} if there exist an orthogonal matrix $ \Omega \in \mathbb{R}^{d \times d}$ and a deterministic vector $ \mu \in \mathbb{R}^{d}$ such that $X \overset{D}{=} \mu + \Omega Z$, where $ Z \in \mathbb{R}^{d}$ is \emph{coordinatewise symmetric} in the sense that $ GZ \overset{ D }{=} Z $ for any diagonal matrix $ G $ with diagonal elements $ G_{jj} \in \{-1, 1\}$.
\end{definition}

\begin{proof}[Proof of Proposition \ref{sec meth: FCS equivalent def}]
On one hand, for a finite Hilbert space say $ \mathbb{R}^{d} $ equipped with the dot product, assume $ X \in \mathbb{R}^{d} $ to satisfy Definition \ref{sec meth: MM3}. For a subset $ \{\psi_{i_{1}}, \dotsc, \psi_{i_{p}}\} $ of an orthonormal basis $ \{\psi_{1}, \dotsc, \psi_{d} \},\ p \leq d $, define the linear operator $ A: \mathbb{R}^{d} \to \mathbb{R}^{p} $ as $ AX = (\langle \psi_{i_{1}}, X \rangle, \dotsc, \langle \psi_{i_{p}}, X \rangle)^{\top} $. In other words, $ A = [\psi_{i_{1}}| \cdots | \psi_{i_{p}}]^{\top} \in \mathbb{R}^{p \times d} $ and in particular $ AA^{\top} = I_d $. By Definition \ref{sec meth: MM3}, we have $ X = \Omega Z + \mu $, where $ \Omega \Omega^{\top} = I_d $ and $ Z $ is coordinatewise symmetric. Now $ (\langle \psi_{i_{1}}, X-\mu \rangle, \dotsc, \langle \psi_{i_{p}}, X -\mu \rangle)^{\top} = A(X-\mu) = A\Omega Z $. Notice that $ A\Omega \Omega^{\top} A^{\top} = I_d $. Therefore, we have $ X $ is FCS. 

On the other hand, by using $ \{ \psi_{i_{1}}, \dotsc, \psi_{i_{d}} \} $, such that $ A = I_d \in  \mathbb{R}^{d \times d} $, it is straightforward to see that if $ X $ is FCS, then it follows Definition \ref{sec meth: MM3}. 
\end{proof}

\begin{lemma} \label{sec proof: lemma 1}
Let $\{ \psi_j(\cdot) \}$ and $\{ \varphi_j(\cdot) \}$ be two orthonormal bases for a Hilbert space $\mathcal{H}$. For any positive integer $p < \infty$, the we have that $G := [\langle \varphi_j, \psi_k \rangle ]_{j=1,\dotsc,p, k = 1, 2, \dotsc}$ satisfy $G G^{\top} = I_p$. 
\end{lemma}

\begin{proof}[Proof of Lemma \ref{sec proof: lemma 1}]
Since $ \{\psi_{j}\} $ and $ \{\varphi_{j}\} $ are both orthonormal basis, we have $ \varphi_{j}(\cdot) = \sum_{k=1}^{\infty} \langle \varphi_{j}, \psi_{k}\rangle \psi_{j}(\cdot)$ and $ \|\varphi_{j}\|^{2} = \sum_{k=1}^{\infty} \langle \varphi_{j}, \psi_{k}\rangle^{2} = 1 $ \citep[Theorem 4.19 on page 59 in][]{Deutsch2012}, which shows that the diagonal of $ G G^{\top} $ are all $ 1$'s. By orthogonality of $ \{\varphi_{j}\} $, we have $ 0 = \langle \varphi_{l}, \varphi_{j} \rangle = \sum_{k=1}^{\infty} \langle \varphi_{l}, \psi_{k} \rangle \langle \varphi_{j}, \psi_{k} \rangle,\ l \neq j$. This shows that the off-diagonal elements in $ GG^{\top} $ are 0's. As a result, $ GG^{\top} = I_p $. 
\end{proof}

\begin{proof}[Proof of Proposition \ref{sec meth: relations between functional distributions}]

(1) $ FE \subset FCS$. On one hand, Proposition 1 in \cite{Boente2014} asserts that if $ X(\cdot) $ is FE distributed and its scatter operator has infinite rank, then $ X(\cdot) $ admits the following decomposition 
\begin{align*}
	X(\cdot) = \mu(\cdot) + SN(\cdot),
\end{align*}
where $N(\cdot)$ is a mean zero Gaussian element and $S$ is
a non-negative random variable independent of $N(\cdot)$. Now for any bounded linear operator $ A: \mathcal{H} \to \mathbb{R}^{d} $ as defined in the proof above for Proposition \ref{sec meth: FCS equivalent def}, $ AN $ is then a mean zero $ d $ dimensional Gaussian vector, which can be decomposed as $ AN = \Omega V $, where $ \Omega $ is the orthogonal matrix, whose columns are the eigenvectors of the covariance matrix of $ AN $; and $ V $ is a mean zero Gaussian vector with diagonal covariance matrix, whose diagonal elements are the corresponding eigenvalues of the covariance matrix of $ AN$. Since $ S $ and $ N(\cdot) $ are independent, we have that $ Z_{A}:=SV $ is coordinatewise symmetric. Thus, $ X $ is FCS. 

On the other hand, it is easy to see that the other direction does not always hold since $ Z $ in Definition 4 only needs to be coordinatewise symmetric, but not necessarily related to Gaussian. Furthermore, individual components of $ Z $ can even come from different distributions, which is prohibited by elliptical distributions. 
	
(2) $ FCS \subset wFCS $. On one hand, if $ X(\cdot) $ is FCS, then for any bounded linear operator $ A: \mathcal{H} \to \mathbb{R}^{d}$ defined in the above proof for Proposition \ref{sec meth: FCS equivalent def}, we have $ A(X-\tilde{X}) = \Omega_{A}(Z_{A} - \tilde{Z}_{A}) $, where $ \tilde{X}(\cdot) $ is an identical copy of $ X(\cdot) $ and $ \Omega_{A} \Omega_{A}^{\top} = I_d $. Since $ Z_{A} $ and $ \tilde{Z}_{A}$ are coordinatewise symmetric, $ Z_{A} - \tilde{Z}_{A} $ is also coordinatewise symmetric. Therefore, $ X(\cdot)-\tilde{X}(\cdot) $ is FCS, and this shows that $ X(\cdot) $ is weakly FCS. 

On the other hand, to see why $ wFCS $ is strictly larger than $ FCS $, consider the example $ X(\cdot) := \sum_{j} \xi_j \psi_j(\cdot)$, where $\xi_j's$ are i.i.d. Chi square distributed and $\{ \psi_j \}$ is an orthornormal basis for $\mathcal{H}$. Then by Lemma \ref{sec proof: lemma 1} above, a straightforward calculation shows that for any $ A: \mathcal{H} \to \mathbb{R}^{d} $ as defined above in the proof for Proposition \ref{sec meth: FCS equivalent def}, $ AX = \Omega_{A} Z_{A} \in \mathbb{R}^{d}$ satisfies that the components of $ Z_{A} $ are independent Chi squares and $ \Omega_{A} \Omega_{A}^{\top} = I_d $. 
Apparently, $ X(\cdot)$ is not FCS because $ Z_{A} $ is not coordinatewise symmetric. However, it can be seen that $ X(\cdot) $ is weakly FCS because with $ \tilde{X}(\cdot) $ as an identical copy of $ X(\cdot) $, $ A(X - \tilde{X}) = \Omega_{A} (Z_{A} - \tilde{Z}_{A})$ and $ Z_{A} - \tilde{Z}_{A} $ is coordinatewise symmetric. 

\end{proof}

\begin{proof}[Proof of Proposition \ref{sec meth: no marginal distribution required}]
Via a typical rescaling argument, without loss of generality, we could assume $ \|\psi\| := \sqrt{\langle \psi, \psi \rangle} = 1 $. Let $ F $ be a well-defined distribution in $ \mathbb{R} $. For any $ \psi(\cdot) \in \mathcal{H} $, such that $ \|\psi\| = 1 $, with the Gram-Schmidt construction \citep[cf. Theorem 8 on page 75 in][]{Cheney2013}, we can add on to $ \psi(\cdot) $ to form an orthonormal basis $ \{\psi_{1}(\cdot) := \psi(\cdot), \psi_{2}(\cdot), \dotsc\} $ for $ \mathcal{H} $. Then we can construct a random function 
\begin{align*}
	X(\cdot) := \sum_{j=1}^{\infty} \xi_{\psi_{j}} \psi_{j}(\cdot),
\end{align*} 
with independent $ \xi_{\psi_{1}} \sim F,\ \xi_{\psi_{2}} \sim F, \dotsc$. Without loss of generality, we assume $ \sum_{j=1}^{\infty} \xi_{\psi_{j}} < \infty $ $ \sum_{j=1}^{\infty} \xi_{\psi_{j}}^{2} < \infty$, so that $ X(\cdot) \in \mathcal{H}$. We can do this because one can always pick a $ q < \infty$ and set $ \xi_{\psi_{q+1}}=0, \xi_{\psi_{q+2}}=0, \dotsc $ when constructing $ X(\cdot) $. Now it is only left to show $ X(\cdot) $ is wFCS. To this end, first let $ \tilde{X}(\cdot) = \sum_{j=1}^{\infty} \tilde{\xi}_{\psi_{j}} \psi_{j}(\cdot) $ be an independent copy of $ X(\cdot) $. Then due to independence, it is straightforward to see that 
\begin{align} \label{sec proof: score difference marginal}
	& (\xi_{\psi_{1}}- \tilde{\xi}_{\psi_{1}}, \dotsc, \xi_{\psi_{d}}- \tilde{\xi}_{\psi_{d}})^{\top}\\
	=&(\langle X-\tilde{X}, \psi_{1} \rangle, \dotsc, \langle X-\tilde{X}, \psi_{d} \rangle)^{\top} \nonumber \\
	=& (\langle X, \psi_{1} \rangle - \langle \tilde{X}, \psi_{1} \rangle, \dotsc, \langle X, \psi_{d} \rangle - \langle \tilde{X}, \psi_{d} \rangle)^{\top},~~~ d \leq \infty \nonumber
\end{align}
is coordinatewise symmetric because marginally $ \langle X, \psi_{i} \rangle - \langle \tilde{X}, \psi_{i} \rangle $ is always symmetric about 0. Next, let $ \{\varphi_{j}(\cdot)\} $ be any orthonormal basis of $ \mathcal{H} $. For any $ p < \infty $, we have
\begin{align*}
	\langle X-\tilde{X}, \varphi_{p} \rangle = \sum_{j=1}^{\infty} (\xi_{\psi_{j}} - \tilde{\xi}_{\psi_{j}}) \langle \psi_{j}, \varphi_{p} \rangle.
\end{align*}

In particular, for any subset $ \{ \varphi_{j_{1}}, \dotsc, \varphi_{j_{p}} \}$ of $ \{\varphi_{j}\} $, we have
\begin{align*}
	& (\langle X-\tilde{X}, \varphi_{j_{1}} \rangle, \dotsc, \langle X-\tilde{X}, \varphi_{j_{p}} \rangle)^{\top} \\
	=& [\langle \psi_{j}, \varphi_{l_{i}} \rangle]_{i = 1, \dotsc, p, j=1, 2, \dotsc } (\xi_{\psi_{1}}-\tilde{\xi}_{\psi_{1}}, \dotsc)^{\top}. 
\end{align*}
By Equation \eqref{sec proof: score difference marginal} above, $ (\xi_{\psi_{1}}-\tilde{\xi}_{\psi_{1}}, \dotsc)^{\top} $ is coordinatewise symmetric, and it can be regarded as the $ Z_{\psi} $ in Definition 4. By Lemma \ref{sec proof: lemma 1} above, we also have $G G^{\top} = I_p$, with $G:= [\langle \psi_{j}, \varphi_{l_{i}} \rangle]_{i = 1, \dotsc, p, j=1, 2, \dotsc }$. Thus $X(\cdot)$ is wFCS.
\end{proof}

\begin{proof}[Proof of Theorem \ref{sec meth thm: Functional Oja}]
Denote the difference between two independent and identically distributed random functions as $D(\cdot) := X(\cdot) - \Tilde{X}(\cdot)$. Then the mean of $D(\cdot)$ is $\mu_D := \mathbb{E} [D(\cdot)] = 0$, and its covariance function is $\Gamma_D(s,t) = 2\Gamma(s,t)$. Let $\phi_{i_1}(\cdot),\ \phi_{i_2}(\cdot), \dotsc$ be the eigenfunctions of $\Gamma(s,t)$ such that their corresponding eigenvalues are nonzero and in descending order $\lambda_{i_1} \geq \lambda_{i_2} \geq \cdots$. The Karhunen-Lo\`eve expansion of $D(\cdot)$ is then $D(\cdot) = \sum_{k \geq 1} \xi_k \phi_{i_k}(\cdot)$, where $\xi_k := \langle D,\ \phi_{i_k} \rangle,\ k \geq 1$, are the scores. The PASS covariance function can then be written as

\begin{align*}
	K(s,t) & = \mathbb{E} \left[ \frac{\sum_{j=1}^{\infty} \xi_j \phi_j(s) \sum_{k=1}^{\infty}\xi_k \phi_k(t)}{\|\sum_{i=1}^{\infty} \xi_i  \phi_i(\cdot)\|^2} \right] \\
	& = \sum_{j, k = 1}^{\infty} \mathbb{E} \left[\frac{\xi_j \xi_k}{ \sum_{i=1}^{\infty} \xi_i^2} \right] \phi_j(s) \phi_k(t)\\
	& = \sum_{j = 1}^{\infty} \mathbb{E} \left[\frac{\xi_j^2}{ \sum_{i=1}^{\infty} \xi_i^2} \right] \phi_j(s) \phi_j(t),
\end{align*}
where the last equation is due to the fact that $ \mathbb{E}  [\xi_j \xi_k / \sum_i \xi_i^2] = 0,\ j \neq k$, which can be shown as follows. For any $ j \neq k $, because $ X(\cdot) $ is weakly FCS, we have with $ d' \geq d \geq \max(j,k) $, $ (\xi_{1}, \dotsc, \xi_{d}) = \Omega_{\phi} Z_{\phi} $, where $ \Omega_{\phi} \in \mathbb{R}^{d \times d'}$ and $ Z_{\phi} \in \mathbb{R}^{d'} $ are defined in Definition 4. In particular $ \Omega_{\phi} \Omega_{\phi}^{\top} = I_d$ and $ (Z_{\phi, 1}, \dotsc, Z_{\phi, j}, \dotsc, Z_{\phi, d'} )^{\top} \overset{D}{=} (Z_{\phi, 1}, \dotsc, -Z_{\phi, j}, \dotsc, Z_{\phi, d'})^{\top}$ for any $ j = 1, \dotsc, d' $. Therefore, 

\begin{align*}
	\mathbb{E} \left[\frac{\xi_j \xi_{k}}{ \sum_{i=1}^{\infty} \xi_i^2} \right] = \left[ \Omega_{\phi} \mathbb{E} \frac{Z_{\phi }Z_{\phi}^{\top}}{\|Z_{\phi}\|^{2}} \Omega_{\phi}^{\top} \right]_{j,k} = 0.
\end{align*}

Let $U_k = \lambda_{i_k}^{-1/2} \xi_k/\sqrt{2} $, then 
\begin{equation*}
	\lambda_{j}(K) := \mathbb{E} \left[ \frac{\xi_{j}^{2}}{\sum_{i=1}^{\infty} \xi_{i}^{2}} \right] = \mathbb{E} \left[ \frac{\lambda_{i_j} U_{j}^{2}}{\sum_{l=1}^{\infty} \lambda_{i_l} U_{l}^{2}} \right].
\end{equation*}

For $ j,\ k \in \mathbb{N} $ (the set of all positive integers) such that $ \lambda_{i_{j}} \leq \lambda_{i_{k}} $, by the exchangeability of $ U_j,\ j = 1,\ 2,\ \dotsc $, we have

\begin{equation*}
 \frac{ \mathbb{E} \left[ \frac{\lambda_{i_j} U_{j}^{2}}{\sum_{l=1}^{\infty} \lambda_{i_l} U_{l}^{2}} \right] }{ \mathbb{E} \left[ \frac{\lambda_{i_k} U_{k}^{2}}{\sum_{l=1}^{\infty} \lambda_{i_l} U_{l}^{2}} \right] }                =              \frac{ \mathbb{E} \left[ \frac{\lambda_{i_j} U_{j}^{2}}{ \lambda_{i_j} U_{j}^{2} + \lambda_{i_k} U_{k}^{2} + R_{j,k}} \right] }{ \mathbb{E} \left[ \frac{\lambda_{i_k} U_{k}^{2}}{\lambda_{i_j} U_{j}^{2} + \lambda_{i_k} U_{k}^{2} + R_{j,k}} \right] }              \leq               \frac{ \mathbb{E} \left[ \frac{\lambda_{i_j} U_{j}^{2}}{ \lambda_{i_j} U_{j}^{2} + \lambda_{i_j} U_{k}^{2} + R_{j,k}} \right] }{ \mathbb{E} \left[ \frac{\lambda_{i_k} U_{k}^{2}}{\lambda_{i_k} U_{j}^{2} + \lambda_{i_k} U_{k}^{2} + R_{j,k}} \right] }                =                  \frac{ \mathbb{E} \left[ \frac{U_{j}^{2}}{U_{j}^{2} + U_{k}^{2} + R_{j,k}/\lambda_{i_{j}}} \right] }{ \mathbb{E} \left[ \frac{U_{j}^{2}}{ U_{j}^{2} + U_{k}^{2} + R_{j,k}/ \lambda_{i_k}} \right] }  \leq 1. 
\end{equation*}

In addition, if $D(\cdot)$ is FE distributed, then the covariance operator of $D(\cdot)$ is proportional to its scatter operator as defined in Definition 3. Without loss of generality, assume that the covariance operator of $D(\cdot)$ is the same as its scatter operator. Define the bounded linear operator $A_d: \mathcal{H} \rightarrow \mathbb{R}^d$ as $A_d D = (\lambda_{i_1}^{-1/2} 1/\sqrt{2}\langle D, \phi_{i_1}\rangle, \dotsc, \lambda_{i_d}^{-1/2} 1/\sqrt{2}\langle D, \phi_{i_d}\rangle)^\top$. Then by Definition 3, ${U}_d := A_dD \sim \epsilon_{d} (0, A_d \mathbf{\Gamma}_D A_d^*, \varphi)$, which is a $ d$-dimensional elliptical distribution. By definition of $A_d$, we have $\langle A_d x, y\rangle  = \sum_{k=1}^d \lambda_{i_k}^{-1/2} y_k \langle x, \phi_{i_k}\rangle = \langle x, 1/\sqrt{2}\sum_{k=1}^d \lambda_{i_k}^{-1/2} y_k \phi_{i_k}\rangle$ for any $ x \in \mathcal{H},\ y \in \mathbb{R}^{d} $. Therefore,
\begin{equation*}
	A_d^*y =1/\sqrt{2} \sum_{k=1}^d \lambda_{i_k}^{-1/2} y_k \phi_{i_k}(\cdot).
\end{equation*}

This implies that $\mathbf{\Gamma}_D A_d^* y = 2 \mathbf{\Gamma} A_d^* y = \sqrt{2} \sum_{k=1}^d \lambda_{i_k}^{1/2} y_k \phi_{i_k}(\cdot)$ (recall $  \Gamma(s,t) = \sum \lambda_j(\Gamma) \phi_j(s) \phi_j(t) $ by Mercer's Lemma). Again by the definition of $A_d$, we have $A_{d}\mathbf{\Gamma}A_d^* y = y$, and this shows that $A_{d}\mathbf{\Gamma}A_d^* = I_d$, which is the identity matrix. This shows that $U_d \sim S(0, \phi),\ \forall d \in \mathbb{N}$, which is a spherical distribution. Then by Lemma 2.1 in \cite{Boente2014}, there is a random vector $w \geq 0$, such that $\forall d \in \mathbb{N}$, $U_d | w \sim N(0, wI_d)$. Let $s = w^{1/2}$, then $U_d \sim s Z_d$, where $Z_d \sim N(0, I_d)$ and $Z_d$ is independent from $s$. Therefore, $U_k \sim sZ_k,\ \forall k \in \mathbb{N}$ with $Z_k$ being i.i.d standard normal random variable and $Z_k$ independent from $s$. This implies that 
\begin{equation*}
    D(\cdot) = \sum_{k\geq1} \lambda_{i_k}^{1/2} \sqrt{2} U_k \phi_{i_k}(\cdot) \sim s \sum_{k\geq1}\lambda_{i_k}^{1/2} \sqrt{2} Z_k \phi_{i_k}(\cdot).
\end{equation*}

As a result, 
\begin{align*}
    K(u, v) & = \mathbb{E} \frac{2s^2 \left(\sum_{k\geq1}\lambda_{i_k}^{1/2} \sqrt{2} Z_k \phi_{i_k}(u) \right) \left(\sum_{k\geq1}\lambda_{i_k}^{1/2} \sqrt{2} Z_k \phi_{i_k}(v) \right)}{2s^2 \|\sum_{k\geq1}\lambda_{i_k}^{1/2} \sqrt{2} Z_k \phi_{i_k}(\cdot)\|^2} \\
    & = \sum_{j\geq1} \left[ \mathbb{E} \frac{\lambda_{i_j} Z_j^2}{\sum_{k\geq1}\lambda_{i_k}Z_k^2}\right]\phi_{i_j}(u) \phi_{i_j}(v).
\end{align*}
This completes the proof.
\end{proof}

\begin{proof}[Proof of Theorem \ref{sec meth thm: eigenratio decay speed}]

In contrast to the proof in \citet*{Durre2016a} that heavily relies on elliptical distribution, we develop a new proof for incorporating the wFCS case. From Equation \eqref{sec meth thm: eigen ratio relation}, the eigenratio can be written as
\begin{equation*}
    \frac{\lambda_j(K)}{\lambda_k(K)} = \frac{\lambda_{j}(\Gamma)}{\lambda_{k}(\Gamma)} \eta,
\end{equation*}
where 
\begin{equation*}
    \eta :=  \mathbb{E} \Big[\frac{U_j^2}{\sum_{i = 1}^\infty \lambda_{i}(\Gamma) U_i^2}\Big]\Big/ \mathbb{E} \Big[\frac{U_k^2}{\sum_{i = 1}^\infty \lambda_{i}(\Gamma) U_i^2}\Big].
\end{equation*}

With this notation, establishing inequality (2.9) is equivalent to showing $\eta \leq 1$. To this end, consider the difference between the numerator and denominator of  $\eta$,
\begin{align}
    & \mathbb{E} \frac{U_j^2}{\sum_{i = 1}^\infty \lambda_{i}(\Gamma) U_i^2} - \mathbb{E} \frac{U_k^2}{\sum_{i = 1}^\infty \lambda_{i}(\Gamma) U_i^2} \label{eq3: eigenratio derivative difference} \\
    = & \mathbb{E} \frac{U_j^2-U_k^2}{\lambda_{j}(\Gamma) U_j^2 + \lambda_{k}(\Gamma) U_k^2 + R_{j,k}} \nonumber \\
    = & \mathbb{E} \frac{U_j^2 - U_k^2}{ \frac{\lambda_{j}(\Gamma) + \lambda_{k}(\Gamma)}{2}(U_j^2 + U_k^2) + \frac{\lambda_{j}(\Gamma) - \lambda_{k}(\Gamma)}{2}(U_j^2 - U_k^2) + R_{j,k}} \nonumber \\
    =& \mathbb{E} \frac{U_j^2 - U_k^2}{ x (U_j^2 + U_k^2) + y (U_j^2 - U_k^2) + R_{j,k}} \nonumber \\
    =:& f(x,y), \nonumber 
\end{align}
where $x = \frac{\lambda_{j}(\Gamma) + \lambda_{k}(\Gamma)}{2}$, $y = \frac{\lambda_{j}(\Gamma) - \lambda_{k}(\Gamma)}{2}$ and $R_{j,k} = \sum_{i \neq j, k}^\infty \lambda_{i}(\Gamma) U_i^2$. Moreover, without loss of generality assume that $\lambda_{j}(\Gamma) > 0$ and $\lambda_{k}(\Gamma) > 0$, thus $x >0$. Let $r:=\lambda_{k}(\Gamma) / \lambda_{j}(\Gamma) $. Then, we have $y = \frac{1-r}{1+r} x$. Thus the difference \eqref{eq3: eigenratio derivative difference} can be regarded as a function of $x \text{ and } r$.,
\begin{equation}
    g(x,r) := \mathbb{E} \frac{U_j^2 - U_k^2}{ x (U_j^2 + U_k^2) + \frac{1-r}{1+r} x (U_j^2 - U_k^2) + R_{j,k}}. \nonumber
\end{equation}

We want to show that $f(x,y) \leq 0$, whenever $\lambda_{j}(\Gamma) \geq \lambda_{k}(\Gamma)$. This is equivalent to show $g(x,r) \leq 0$, whenever $0 < r \leq 1$. This can be seen by taking partial derivative with respect to $r$,
\begin{align*}
    \frac{\partial g(x,r)}{\partial r} & = \mathbb{E} \frac{(U_j^2-U_k^2)^2 \frac{2}{(1+r)^2}}{[x (U_j^2 + U_k^2) + \frac{1-r}{1+r} x (U_j^2 - U_k^2) + R_{1,2}]^2},
\end{align*}
which shows that $\frac{\partial g(x,r)}{\partial r} > 0,\ \forall r$. Notice that $g(x, 1) = 0$. As a result, $g(x,r)$ decreases from $0$ as $r$ decreases from $1$. This shows that $g(x,r) \leq 0,\ \forall 0 < r \leq 1$.

\end{proof}

\begin{proof}[Proof of Corollary \ref{sec meth cor: eigen ratio relation}]
	For any $ Q \geq 1 $, we have 
	\begin{align*}
		\frac{\sum_{i=1}^{Q} \lambda_{i}(K)}{\sum_{i=1}^{\infty} \lambda_{i}(K)} = \frac{\sum_{i=1}^{Q} \lambda_{i}(K)}{\sum_{i=1}^{Q} \lambda_{i}(K) +  \sum_{i=Q+1}^{\infty} \lambda_{i}(K)} = \frac{1}{1 + \frac{\sum_{i=Q+1}^{\infty}\lambda_{i}(K)}{\sum_{i=1}^{Q}\lambda_{i}(K)}},
	\end{align*}
	and similarly
	\begin{align*}
		\frac{\sum_{i=1}^{Q} \lambda_{i}(\Gamma)}{\sum_{i=1}^{\infty} \lambda_{i}(\Gamma)} = \frac{1}{1 + \frac{\sum_{i=Q+1}^{\infty}\lambda_{i}(\Gamma)}{\sum_{i=1}^{Q}\lambda_{i}(\Gamma)}}.
	\end{align*}
	
	Now observe that
	\begin{align*}
		\frac{\sum_{i=Q+1}^{\infty}\lambda_{i}(\Gamma)}{\sum_{i=1}^{Q}\lambda_{i}(\Gamma)} = \frac{\sum_{i=Q+1}^{\infty}\lambda_{i}(\Gamma)}{\sum_{i=1}^{Q-1}\lambda_{i}(\Gamma)+\lambda_{Q}(\Gamma)} = \frac{\sum_{i=Q+1}^{\infty}\frac{\lambda_{i}(\Gamma)}{\lambda_{Q}(\Gamma)}}{\sum_{i=1}^{Q-1} \frac{\lambda_{i}(\Gamma)}{\lambda_{Q}(\Gamma)} +1},
	\end{align*}
	\noindent and 
	\begin{align*}
		\frac{\sum_{i=Q+1}^{\infty}\lambda_{i}(K)}{\sum_{i=1}^{Q}\lambda_{i}(K)} = \frac{\sum_{i=Q+1}^{\infty}\frac{\lambda_{i}(K)}{\lambda_{Q}(K)}}{\sum_{i=1}^{Q-1} \frac{\lambda_{i}(K)}{\lambda_{Q}(K)} +1}.
	\end{align*}
	
	By Theorem \ref{sec meth thm: eigenratio decay speed},
	\begin{align*}
		\frac{\lambda_{i}(K)}{\lambda_{Q}(K)} \leq \frac{\lambda_{i}(\Gamma)}{\lambda_{Q}(\Gamma)},\ \forall i \leq Q,
	\end{align*}
	and
	\begin{align*}
		\frac{\lambda_{i}(K)}{\lambda_{Q}(K)} \geq \frac{\lambda_{i}(\Gamma)}{\lambda_{Q}(\Gamma)},\ \forall i \geq Q.
	\end{align*}
	Therefore,
	\begin{align*}
		\frac{\sum_{i=Q+1}^{\infty}\lambda_{i}(K)}{\sum_{i=1}^{Q}\lambda_{i}(K)} \geq \frac{\sum_{i=Q+1}^{\infty}\lambda_{i}(\Gamma)}{\sum_{i=1}^{Q}\lambda_{i}(\Gamma)},
	\end{align*}
	which yields
	\begin{align*}
		\frac{\sum_{i=1}^{Q} \lambda_{i}(K)}{\sum_{i=1}^{\infty} \lambda_{i}(K)} \leq \frac{\sum_{i=1}^{Q} \lambda_{i}(\Gamma)}{\sum_{i=1}^{\infty} \lambda_{i}(\Gamma)}.
	\end{align*}
This completes the proof.
\end{proof}

\begin{proof}[Proof of Corollary \ref{sec meth thm: functional Oja with noise}]
	The norm $ \|Y_{p}(\cdot) -\widetilde{Y}_{p}(\cdot) \|^{2} $ on the denominator of $K^*(t_j, t_k)$, which is defined by Equation \eqref{sec meth: PASS with noise def} in the main paper, can be written as
	\begin{align*}
		\|Y_{p}(\cdot) -\widetilde{Y}_{p}(\cdot) \|^{2} & = \| X_{p}(\cdot) - \widetilde{X}_{p}(\cdot) + \epsilon_{p}(\cdot) - \widetilde{\epsilon}_{p}(\cdot) \|^{2} \\
		& = \| X_{p}(\cdot) - \widetilde{X}_{p}(\cdot) \|^{2} + \| \epsilon_{p}(\cdot) - \widetilde{\epsilon}_{p}(\cdot) \|^{2} \\
		& + 2 \langle X_{p}(\cdot) - \widetilde{X}_{p}(\cdot),\  \epsilon_{p}(\cdot) - \widetilde{\epsilon}_{p}(\cdot)\rangle, 
	\end{align*}
	where the last term is
	\begin{align*}
		\langle X_{p}(\cdot) - \widetilde{X}_{p}(\cdot),\  \epsilon_{p}(\cdot) - \widetilde{\epsilon}_{p}(\cdot)\rangle & = \frac{1}{N} \sum_{i = 1}^{N} [X(t_{i}) - \widetilde{X}(t_{i})] [\epsilon(t_{i}) - \widetilde{\epsilon}(t_{i})]. 
	\end{align*}
	Since $ X(t_{i}) - \widetilde{X}(t_{i}) $ and $ \epsilon(t_{i}) - \widetilde{\epsilon}(t_{i}) $ are independent, it is straightforward to see that
	\begin{align*}
		\mathbb{E} \left\{ \frac{1}{N} \sum_{i = 1}^{N} [X(t_{i}) - \widetilde{X}(t_{i})] [\epsilon(t_{i}) - \widetilde{\epsilon}(t_{i})] \right\}= 0
	\end{align*}
	and 
	\begin{align*}
		{\rm Var} \left\{ \frac{1}{N} \sum_{i = 1}^{N} [X(t_{i}) - \widetilde{X}(t_{i})] [\epsilon(t_{i}) - \widetilde{\epsilon}(t_{i})] \right\} \leq \frac{4\sigma^{2}}{N} \underset{1 \leq j \leq N}{\max} \Gamma(t_{j}, t_{j}). 
	\end{align*}
	Then by weak law of large numbers, as $ N \rightarrow \infty $, we have $\langle X_{p}(\cdot) - \widetilde{X}_{p}(\cdot),\  \epsilon_{p}(\cdot) - \widetilde{\epsilon}_{p}(\cdot)\rangle  = o_{p}(1).$

In addition $ \| X_{p}(\cdot) - \widetilde{X}_{p}(\cdot) \|^{2}  = \frac{1}{N} \sum_{k=1}^{N} [X(t_{k}) - \widetilde{X}(t_{k})]^{2} $ is the Riemann sum of $ [X(t_{k}) - \widetilde{X}(t_{k})]^{2} $, thus $ \| X_{p}(\cdot) - \widetilde{X}_{p}(\cdot) \|^{2}  \rightarrow_{p} \int_{0}^{1} [X(t) - \widetilde{X}(t)]^{2}dt = \|X(\cdot) - \widetilde{X}(\cdot)\|^{2}$. Now write $\|Y_p(\cdot) - \tilde{Y}_p(\cdot) \|^2$ as the following:
\begin{align*}
    \|Y_{p}(\cdot) -\widetilde{Y}_{p}(\cdot) \|^{2} & = \| X(\cdot) - \widetilde{X}(\cdot) \|^{2} + \| \epsilon_{p}(\cdot) - \widetilde{\epsilon}_{p}(\cdot) \|^{2} + R_{X, \epsilon},
\end{align*}
where $R_{X, \epsilon} := \|X_p(\cdot)-\tilde{X}_p(\cdot) \|^2 - \|X(\cdot)-\tilde{X}(\cdot) \|^2 + 2 \langle X_{p}(\cdot) - \widetilde{X}_{p}(\cdot),\  \epsilon_{p}(\cdot) - \widetilde{\epsilon}_{p}(\cdot)\rangle = o_p(1)$. Then
\begin{align*}
     & \frac{[Y(t_{j})-\widetilde{Y}(t_{j})][Y(t_{l})-\widetilde{Y}(t_{l})]}{\| Y_{p}(\cdot)-\widetilde{Y}_{p}(\cdot) \|^{2}} \\
     =&  \frac{[Y(t_{j})-\widetilde{Y}(t_{j})][Y(t_{l})-\widetilde{Y}(t_{l})]}{\| X(\cdot) - \widetilde{X}(\cdot) \|^{2} + \| \epsilon_{p}(\cdot) - \widetilde{\epsilon}_{p}(\cdot) \|^{2} + R_{X, \epsilon}} \\
     =& \left\{\frac{[Y(t_{j})-\widetilde{Y}(t_{j})][Y(t_{l})-\widetilde{Y}(t_{l})]}{\| X(\cdot) - \widetilde{X}(\cdot) \|^{2} + \| \epsilon_{p}(\cdot) - \widetilde{\epsilon}_{p}(\cdot) \|^{2}} \right\} \Bigg/ \left\{1 + \frac{R_{X,\epsilon}}{\| X(\cdot) - \widetilde{X}(\cdot) \|^{2} + \| \epsilon_{p}(\cdot) - \widetilde{\epsilon}_{p}(\cdot) \|^{2}}\right\} \\
     =& \frac{[Y(t_{j})-\widetilde{Y}(t_{j})][Y(t_{l})-\widetilde{Y}(t_{l})]}{\| X(\cdot) - \widetilde{X}(\cdot) \|^{2} + \| \epsilon_{p}(\cdot) - \widetilde{\epsilon}_{p}(\cdot) \|^{2}} \times \left\{ 1 - \zeta \right\},
\end{align*}
where $|\zeta| \leq |R_{X,\epsilon}/\{\| X(\cdot) - \widetilde{X}(\cdot) \|^{2} + \| \epsilon_{p}(\cdot) - \widetilde{\epsilon}_{p}(\cdot) \|^{2}\}|$. Observe that with 
\begin{align*}
    & \| X(\cdot) - \widetilde{X}(\cdot) \|^{2} + \| \epsilon_{p}(\cdot) - \widetilde{\epsilon}_{p}(\cdot) \|^{2}\\
    =& \| X(\cdot) - \widetilde{X}(\cdot) \|^{2} - \| X_p(\cdot) - \widetilde{X}_p(\cdot) \|^{2} + \| X_p(\cdot) - \widetilde{X}_p(\cdot) \|^{2} + \| \epsilon_{p}(\cdot) - \widetilde{\epsilon}_{p}(\cdot) \|^{2},
\end{align*}
we have
\begin{align*}
    & \left|\frac{R_{X,\epsilon}}{\| X(\cdot) - \widetilde{X}(\cdot) \|^{2} + \| \epsilon_{p}(\cdot) - \widetilde{\epsilon}_{p}(\cdot) \|^{2}} \right| \\
    =&\left| \frac{\|X(\cdot)-\tilde{X}(\cdot) \|^2 - \|X_p(\cdot)-\tilde{X}_p(\cdot) \|^2 - 2 \langle X_{p}(\cdot) - \widetilde{X}_{p}(\cdot),\  \epsilon_{p}(\cdot) - \widetilde{\epsilon}_{p}(\cdot)\rangle}{\| X(\cdot) - \widetilde{X}(\cdot) \|^{2} - \| X_p(\cdot) - \widetilde{X}_p(\cdot) \|^{2} + \| X_p(\cdot) - \widetilde{X}_p(\cdot) \|^{2} + \| \epsilon_{p}(\cdot) - \widetilde{\epsilon}_{p}(\cdot) \|^{2}} \right| \\
    \leq& 1.
\end{align*}
Therefore,
\begin{align}
    & \left| \frac{[Y(t_{j})-\widetilde{Y}(t_{j})][Y(t_{l})-\widetilde{Y}(t_{l})]}{\| X(\cdot) - \widetilde{X}(\cdot) \|^{2} + \| \epsilon_{p}(\cdot) - \widetilde{\epsilon}_{p}(\cdot) \|^{2}} \times \zeta \right| \nonumber\\
    \leq &\left| \frac{[Y(t_{j})-\widetilde{Y}(t_{j})][Y(t_{l})-\widetilde{Y}(t_{l})]}{\| X(\cdot) - \widetilde{X}(\cdot) \|^{2} + \| \epsilon_{p}(\cdot) - \widetilde{\epsilon}_{p}(\cdot) \|^{2}} \times \frac{R_{X,\epsilon}}{\| X(\cdot) - \widetilde{X}(\cdot) \|^{2} + \| \epsilon_{p}(\cdot) - \widetilde{\epsilon}_{p}(\cdot) \|^{2}} \right| \label{sec proof: slutsky thm res}\\
    \leq& \left| \frac{[Y(t_{j})-\widetilde{Y}(t_{j})][Y(t_{l})-\widetilde{Y}(t_{l})]}{\| X(\cdot) - \widetilde{X}(\cdot) \|^{2} + \| \epsilon_{p}(\cdot) - \widetilde{\epsilon}_{p}(\cdot) \|^{2}} \right| \nonumber \\
    \leq & \left| \frac{[Y(t_{j})-\widetilde{Y}(t_{j})][Y(t_{l})-\widetilde{Y}(t_{l})]}{\| X(\cdot) - \widetilde{X}(\cdot) \|^{2}} \right|. \nonumber
\end{align}
Recall that $R_{X,\epsilon} = o_p(1)$, then Slutsky's Theorem gives that \eqref{sec proof: slutsky thm res} above is $o_p(1)$. Then the above first inequality gives that
\begin{align*}
    \left| \frac{[Y(t_{j})-\widetilde{Y}(t_{j})][Y(t_{l})-\widetilde{Y}(t_{l})]}{\| X(\cdot) - \widetilde{X}(\cdot) \|^{2} + \| \epsilon_{p}(\cdot) - \widetilde{\epsilon}_{p}(\cdot) \|^{2}} \times \zeta \right| = o_p(1).
\end{align*}
Further, with $\epsilon(t_j)$ being independent from $X(\cdot)$, the assumption $\mathbb{E}\{\|X(\cdot) - \tilde{X}(\cdot)\|^{-2} \} < \infty$ implies that 
\begin{align*}
    \mathbb{E} \left| \frac{[Y(t_{j})-\widetilde{Y}(t_{j})][Y(t_{l})-\widetilde{Y}(t_{l})]}{\| X(\cdot) - \widetilde{X}(\cdot) \|^{2}} \right| < \infty.
\end{align*}
Then the Dominated Convergence Theorem gives
\begin{align*}
    \mathbb{E} \left| \frac{[Y(t_{j})-\widetilde{Y}(t_{j})][Y(t_{l})-\widetilde{Y}(t_{l})]}{\| X(\cdot) - \widetilde{X}(\cdot) \|^{2} + \| \epsilon_{p}(\cdot) - \widetilde{\epsilon}_{p}(\cdot) \|^{2}} \times \zeta \right| = o(1).
\end{align*}

	As a result,
	\begin{align*}
		K^{*}(t_{j},t_{l})  =& \mathbb{E} \frac{[Y(t_{j})-\widetilde{Y}(t_{j})][Y(t_{l})-\widetilde{Y}(t_{l})]}{\| Y_{p}(\cdot)-\widetilde{Y}_{p}(\cdot) \|^{2}} \\
		=&\mathbb{E} \frac{[Y(t_{j})-\widetilde{Y}(t_{j})][Y(t_{l})-\widetilde{Y}(t_{l})]}{\| X(\cdot) - \widetilde{X}(\cdot) \|^{2} + \| \epsilon_{p}(\cdot) - \widetilde{\epsilon}_{p}(\cdot) \|^{2}} + o(1) \\
		=& \mathbb{E} \frac{[X({t_{j}})-\widetilde{X}(t_{j})][X(t_{l})-\widetilde{X}(t_{l})]}{\| X(\cdot) - \widetilde{X}(\cdot) \|^{2} + \| \epsilon_{p}(\cdot) - \widetilde{\epsilon}_{p}(\cdot) \|^{2}} \\
		& + \mathbb{E} \frac{[X(t_{j})-\widetilde{X}(t_{j})][\epsilon(t_{l})-\widetilde{\epsilon}(t_{l})]}{\| X(\cdot) - \widetilde{X}(\cdot) \|^{2} + \| \epsilon_{p}(\cdot) - \widetilde{\epsilon}_{p}(\cdot) \|^{2}} + \mathbb{E} \frac{[X(t_{l})-\widetilde{X}(t_{l})][\epsilon(t_{j})-\widetilde{\epsilon}(t_{j})]}{\| X(\cdot) - \widetilde{X}(\cdot) \|^{2} + \| \epsilon_{p}(\cdot) - \widetilde{\epsilon}_{p}(\cdot) \|^{2}} \\
		& + \mathbb{E} \frac{[\epsilon(t_{j})-\widetilde{\epsilon}(t_{j})][\epsilon(t_{l})-\widetilde{\epsilon}(t_{l})]}{\| X(\cdot) - \widetilde{X}(\cdot) \|^{2} + \| \epsilon_{p}(\cdot) - \widetilde{\epsilon}_{p}(\cdot) \|^{2}} + o(1).
	\end{align*}
	
	Since $ \epsilon(t_{1}), \dotsc, \epsilon(t_{N}) $ are independent, it is straightforward to see that $ [\epsilon(t_{1}) -\tilde{\epsilon}(t_{1}) , \dotsc, \epsilon(t_{N}) -\tilde{\epsilon}(t_{N}) ] $ is coordinatewise symmetric as defined in Definition \ref{sec meth: MM3}. In addition, by the assumption that $ X(\cdot) - \widetilde{X}(\cdot) $ and $ \epsilon(t_{i}) - \widetilde{\epsilon}(t_{i}) $ are independent, we have 
	
	\begin{equation*}
		\mathbb{E} \frac{[X(t_{l})-\widetilde{X}(t_{l})][\epsilon(t_{j})-\widetilde{\epsilon}(t_{j})]}{\| X(\cdot) - \tilde{X}(\cdot) \|^{2} + \| \epsilon_{p}(\cdot) - \widetilde{\epsilon}_{p}(\cdot) \|^{2}} = - \mathbb{E} \frac{[X(t_{l})-\widetilde{X}(t_{l})][\epsilon(t_{j})-\widetilde{\epsilon}(t_{j})]}{\| X(\cdot) - \tilde{X}(\cdot) \|^{2} + \| \epsilon_{p}(\cdot) - \widetilde{\epsilon}_{p}(\cdot) \|^{2}},
	\end{equation*}
	which implies that 
	\begin{equation*}
		\mathbb{E} \frac{[X(t_{l})-\widetilde{X}(t_{l})][\epsilon(t_{j})-\widetilde{\epsilon}(t_{j})]}{\| X(\cdot) - \tilde{X}(\cdot) \|^{2} + \| \epsilon_{p}(\cdot) - \widetilde{\epsilon}_{p}(\cdot) \|^{2}} = 0.
	\end{equation*}

	Similarly, we have 
	 
	 \begin{equation*}
	 	\mathbb{E} \frac{[X(t_{j})-\widetilde{X}(t_{j})][\epsilon(t_{l})-\widetilde{\epsilon}(t_{l})]}{\| X(\cdot) - \tilde{X}(\cdot) \|^{2} + \| \epsilon_{p}(\cdot) - \widetilde{\epsilon}_{p}(\cdot) \|^{2}} = 0.
	 \end{equation*}
	
	Thus, we have  
	\begin{align*}
	K^{*}(t_{j},t_{l})  &= \mathbb{E} \frac{[X({t_{j}})-\widetilde{X}(t_{j})][X(t_{l})-\widetilde{X}(t_{l})]}{\| X(\cdot) - \tilde{X}(\cdot) \|^{2} + \| \epsilon_{p}(\cdot) - \widetilde{\epsilon}_{p}(\cdot) \|^{2}} \\
	& ~~~+ \mathbb{E} \frac{[\epsilon(t_{j})-\widetilde{\epsilon}(t_{j})][\epsilon(t_{l})-\widetilde{\epsilon}(t_{l})]}{\| X(\cdot) - \tilde{X}(\cdot) \|^{2} + \| \epsilon_{p}(\cdot) - \widetilde{\epsilon}_{p}(\cdot) \|^{2}}+ o(1).
	\end{align*}
	Similar to the argument for Theorem \ref{sec meth thm: Functional Oja}, the first part of $ K^{*}(t_{j}, t_{l}) $ can be written as
	\begin{align*}
		& \mathbb{E} \frac{[X(t_{j})-\widetilde{X}(t_{j})][X(t_{l})-\widetilde{X}(t_{l})]}{\| X(\cdot) - \widetilde{X}(\cdot) \|^{2} + \| \epsilon_{p}(\cdot) - \widetilde{\epsilon}_{p}(\cdot) \|^{2}} +o(1)\\
		 = & \sum_{j=1}^{\infty} \mathbb{E} \frac{(\xi_{j}-\tilde{\xi}_{j})^{2}}{\sum_{l=1}^\infty (\xi_l -\tilde{\xi}_l)^2 + \| \epsilon_{p}(\cdot) - \widetilde{\epsilon}_{p}(\cdot) \|^{2}} \phi_{j}(t_{l}) \phi_{j}(t_{j}) +o(1),
	\end{align*}
	where $ \xi_{j} := \langle X_{j},\ \phi_{j}\rangle  $ and $ \tilde{\xi}_{j} := \langle \widetilde{X},\ \phi_{j}\rangle$. Letting $ U_{j} = \lambda_{j}^{-1/2} (\xi_{j} - \tilde{\xi}_{j}) $, we then have 
	\begin{align*}
		\lambda_j(K^{*}) :=& \mathbb{E} \frac{(\xi_{j}-\tilde{\xi}_{j})^{2}}{\sum_{l=1}^\infty (\xi_l -\tilde{\xi}_l)^2 + \| \epsilon_{p}(\cdot) - \widetilde{\epsilon}_{p}(\cdot) \|^{2}}\\
		=&\mathbb{E} \frac{ \lambda_{j} U_{j}^{2} }{\sum_{i=1}^{\infty} \lambda_{i} U_{i}^{2} + \| \epsilon_{p}(\cdot) - \widetilde{\epsilon}_{p}(\cdot) \|^{2}}.
	\end{align*}
	Then by using similar arguments in Theorem \ref{sec meth thm: Functional Oja}, we have $ \lambda_j(K^{*}) \leq \lambda_k(K^{*}) $ whenever $ \lambda_j(\Gamma) \leq \lambda_k(\Gamma) $ for $ j,\ k \in \mathbb{N} $.
	
	For the error term, the second part of $ K^{*}(t_{j}, t_{l}) $, by the coordinatewise symmetry, for any $ j \neq l $, we have
	
	\begin{align*}
		\mathbb{E} \frac{[\epsilon(t_{j})-\widetilde{\epsilon}(t_{j})][\epsilon(t_{l})-\widetilde{\epsilon}(t_{l})]}{\| X(\cdot) - \tilde{X}(\cdot) \|^{2} + \| \epsilon_{p}(\cdot) - \widetilde{\epsilon}_{p}(\cdot) \|^{2}} = \mathbb{E} \frac{[\epsilon(t_{j})-\widetilde{\epsilon}(t_{j})][-\epsilon(t_{l})+\widetilde{\epsilon}(t_{l})]}{\| X(\cdot) - \tilde{X}(\cdot) \|^{2} + \| \epsilon_{p}(\cdot) - \widetilde{\epsilon}_{p}(\cdot) \|^{2}}.
	\end{align*}
	Due to the above equation plus $\epsilon(t_j)$'s are independent and identically distributed, we have
	\begin{align*}
		\mathbb{E} \frac{[\epsilon(t_{j})-\widetilde{\epsilon}(t_{j})][\epsilon(t_{l})-\widetilde{\epsilon}(t_{l})]}{\| X(\cdot) - \widetilde{X}(\cdot) \|^{2} + \| \epsilon_{p}(\cdot) - \widetilde{\epsilon}_{p}(\cdot) \|^{2}}  = C 1 (j=l),
	\end{align*}
	where
	\begin{align*}
	    C := \mathbb{E} \frac{[\epsilon(t_{j})-\widetilde{\epsilon}(t_{j})]^{2}}{\| X(\cdot) - \widetilde{X}(\cdot) \|^{2} + \| \epsilon_{p}(\cdot) - \widetilde{\epsilon}_{p}(\cdot) \|^{2}}.
	\end{align*}
This completes the proof.	
\end{proof}

\section{Appendix B}\label{supp: algorithm convergence}

Here we provide some discussions on the convergence of Algorithms 1 given in the main paper. Observe that with a large sample of $ \{V_{ij,l}\} $, the $ f_{k}(\Lambda^{(a)})$'s in Algorithm 1 are close to their population counterparts $ \mathbb{E} [f_{k}(\Lambda^{(a)})] $ by the law of large numbers. Therefore, without loss of generality, we focus on the convergence of iterating the functions $ g(x) = (g_{1}(x), \dotsc, g_{Q-1}(x))^{\top}$, where $ x = (x_{1}, \dotsc, x_{Q-1})^{\top},\ x_{l} \in [0,1],\ l = 1, \dotsc, Q-1 $, and
\begin{align*}
	g_{i}(x) = \frac{\lambda_{i+1}(K)}{\lambda_{1}(K)} \frac{\mathbb{E} \frac{ U_{1}^2}{U_{1}^2 + \sum_{l=1}^{Q-1} x_{l} U_{l+1}^2}}{\mathbb{E} \frac{ U_{i+1}^2}{U_{1}^2 + \sum_{l=1}^{Q-1} x_{l} U_{l+1}^2}},\ i = 1, \dotsc, Q-1.
\end{align*}

Recall these functions are those defined by Equations (12) in the main paper, and $ f_{k}(\Lambda^{(a)})$'s are used to estimate $ \mathbb{E}[U_{i}^{2}/(U_{i}^{2}+\sum x_{l}^{(a)} U_{l+1}^{2})] $ in the algorithms.

Let $ x^{*} = (x_{1}^{*}, \dotsc, x_{Q-1}^{*})^{\top} $ denote a fixed point of $ g(x) $. The following proposition shows that with a sufficiently close initial value $ x^{0} $ and under a relative weak condition, the iterative algorithm $ x^{a+1} = g(x^{a}) $ converges to the fixed point $ x^{*} $.

\begin{proposition} \label{sec meth: convergence proposition}
	If $ U_{l},\ l = 1, \dotsc, Q$ satisfy 
	\begin{align} \label{sec meth: convergence condition}
		\sum_{l=1}^{Q-1} \left| -\frac{ \mathbb{E} \frac{U_{1}^{2}U_{l+1}^{2}}{ (U_{1}^{2} + \sum_{i}^{Q-1} x_{i}^{*} U_{i+1}^{2})^{2}} }{ \mathbb{E} \frac{U_{ 1}^{2}}{U_{1}^{2} + \sum_{i}^{Q-1} x_{i}^{*} U_{ i+1}^{2}}} +  \frac{ \mathbb{E} \frac{U_{k+1}^{2}U_{ l+1}^{2}}{(U_{1}^{2} + \sum_{i}^{Q-1} x_{i}^{*} U_{ i+1}^{2})^{2}}}{ \mathbb{E} \frac{U_{ k+1}^{2}}{U_{1}^{2} + \sum_{i}^{Q-1} x_{i}^{*} U_{ i+1}^{2}}} \right| < \frac{1}{x_{k}^{*}},
	\end{align}
	
	\noindent then starting from $ x^{0} $ such that $\|x^{0} - x^{*}\|_{\infty} := \underset{i=1, \dotsc, Q-1}{\max} |x_{i}^{0} - x_{i}^{*}| \leq d $ for a small $ d $, the iteration $ x^{a+1} = g(x^{a}) $ converges to the fixed point $ x^{*} $. 
\end{proposition}

A natural question following Proposition \ref{sec meth: convergence proposition} is how strong the condition \eqref{sec meth: convergence condition} is and how much it hinders the applicability of Algorithms 1 and 2 in practice. Without knowledge of the distribution of $ \{U_{j,l}\} $, it is almost impossible to give a theoretical judgment on the strictness of this condition. In order not to distract the main message of this paper, we defer this theoretical exposition to future projects. Instead, we conducted a simulation study with a large sample of $ \{U_{j,l}\} $ generated from various distributions and with randomly selected $ x^{*} $ to estimate the left hand side of Inequality \eqref{sec meth: convergence condition}. The sample version of Inequality \eqref{sec meth: convergence condition} prevailed in all of our simulation scenarios. In addition, in our simulation study for estimating the eigen components that is presented in Section 5 in the main paper, Algorithms 1 never failed to converge.

\begin{proof}[Proof of Proposition \ref{sec meth: convergence proposition}]
		This proof follows the arguments in Section 8.6.1 from \cite{Lambers2018}. The goal is to show the mapping defined by function $ g(x) $ is a contraction in a small neighborhood of the fixed point, and this is achieved by bounding the sup-norm induced operator norm of its Jacobin matrix denoted by $ \|J_{g}(x^{*})\|_{\infty} := \underset{}{\max} \sum_{l} | J_{g, kl}(x^{*})| $, where
		
		\begin{align*}
			J_{g}(\cdot) := \left[ \frac{\partial g_{k}}{\partial x_{l}} (\cdot) \right]_{k, l = 1, \dotsc, Q-1}.
		\end{align*}
		
		Then the contraction mapping theorem  \citep[cf. Theorem 1.1 in][]{Berinde2007} is invoked to finish the proof. First, we obtain the partial derivatives through the following standard calculation:
		\begin{align*}
			\frac{\partial g_{k}}{\partial x_{l}} (x^{*})& = \frac{\lambda_{k+1}}{\lambda_{1}} 
			\left[ \frac{\mathbb{E} \frac{U_{  k+1}^{2}}{U_{ 1}^{2} + \sum_{i}^{Q-1} {x}_{i}^{*} U_{  i+1}^{2}} \mathbb{E} \frac{-U_{ 1}^{2}U_{  l+1}^{2}}{ (U_{ 1}^{2} + \sum_{i}^{Q-1} x_{i}^{*} U_{  i+1}^{2})^{2}} }{ \left(\mathbb{E} \frac{U_{  k+1}^{2}}{U_{ 1}^{2} + \sum_{i}^{Q-1} x_{i}^{*} U_{  i+1}^{2}}\right)^{2}} \right. \\
			& \hspace{0.6in} \left. +  \frac{\mathbb{E} \frac{U_{  1}^{2}}{U_{ 1}^{2} + \sum_{i}^{Q-1} x_{i}^{*} U_{  i+1}^{2}} \mathbb{E} \frac{U_{ k+1}^{2}U_{  l+1}^{2}}{(U_{ 1}^{2} + \sum_{i}^{Q-1} x_{i}^{*} U_{  i+1}^{2})^{2}}}{ \left(\mathbb{E} \frac{U_{  k+1}^{2}}{U_{ 1}^{2} + \sum_{i}^{Q-1} x_{i}^{*} U_{  i+1}^{2}}\right)^{2}}\right] \\
			&= - 
			\frac{ x_{k}^{*}\mathbb{E} \frac{U_{ 1}^{2}U_{  l+1}^{2}}{ (U_{ 1}^{2} + \sum_{i}^{Q-1} x_{i}^{*} U_{  i+1}^{2})^{2}} }{ \mathbb{E} \frac{U_{  1}^{2}}{U_{ 1}^{2} + \sum_{i}^{Q-1} x_{i}^{*} U_{  i+1}^{2}}} +  \frac{ x_{k}^{*}\mathbb{E} \frac{U_{ k+1}^{2}U_{  l+1}^{2}}{(U_{ 1}^{2} + \sum_{i}^{Q-1} x_{i}^{*} U_{  i+1}^{2})^{2}}}{ \mathbb{E} \frac{U_{  k+1}^{2}}{U_{ 1}^{2} + \sum_{i}^{Q-1} x_{i}^{*} U_{  i+1}^{2}}},
		\end{align*}
		where the last equation is because at the fixed point $ x^{*} $, we have
		\begin{align*}
			x_{k}^{*} = \frac{\lambda_{k+1}}{\lambda_{1}} \frac{\mathbb{E} \frac{U_{ 1}^{2}}{ U_{ 1}^{2} + \sum_{i}^{Q-1} x_{i}^{*} U_{  i+1}^{2}}}{\mathbb{E} \frac{U_{ k+1}^{2}}{ U_{ 1}^{2} + \sum_{i}^{Q-1} x_{i}^{*} U_{  i+1}^{2}}}. 
		\end{align*}
		
		Then for all $ k = 1, \dotsc, Q-1 $, we have
		
		\begin{align*}
			& \sum_{l=1}^{Q-1}\left| \frac{\partial g_{k} }{\partial x_{l}}(x^{*}) \right| \\
			= & \sum_{l=1}^{Q-1} \left| -\frac{ x_{k}^{*}\mathbb{E} \frac{U_{ 1}^{2}U_{  l+1}^{2}}{ (U_{ 1}^{2} + \sum_{i}^{Q-1} x_{i}^{*} U_{  i+1}^{2})^{2}} }{ \mathbb{E} \frac{U_{  1}^{2}}{U_{ 1}^{2} + \sum_{i}^{Q-1} x_{i}^{*} U_{  i+1}^{2}}} +  \frac{ x_{k}^{*}\mathbb{E} \frac{U_{ k+1}^{2}U_{  l+1}^{2}}{(U_{ 1}^{2} + \sum_{i}^{Q-1} x_{i}^{*} U_{  i+1}^{2})^{2}}}{ \mathbb{E} \frac{U_{  k+1}^{2}}{U_{ 1}^{2} + \sum_{i}^{Q-1} x_{i}^{*} U_{  i+1}^{2}}} \right|\\
			< & x_{k}^{*} \frac{1}{x_{k}^{*}} = 1. \tag{Due to condition \ref{sec meth: convergence condition}}
		\end{align*}
		
		As a result, 
		
		\begin{align*}
			\left\| \frac{\partial g_{k}}{\partial x_{l}} (x^{*}) \right\|_{\infty}  =  \underset{k}{\max} \sum_{l=1}^{Q-1}\left| \frac{\partial g_{k} }{\partial x_{l}}(x^{*}) \right| < 1. 
		\end{align*}
		
		Since $ g(x) $ is continuously differentiable under regularity conditions, let $ N_{d} := \{x: \|x- x^{*}\|_{\infty} \leq d \} $ be a sufficiently small neighborhood of $ x^{*} $ such that $ L:=\sup_{z \in N_{d}} \| \partial g_{k} / \partial g_{l} (z) \|_{\infty} < 1 $. Then by mean value inequality (cf. Corollary 11.34 in \citeauthor{Wade2010}, \citeyear{Wade2010}), we have that 
		
		\begin{align*}
			\|g(x) - x^{*} \|_{\infty} \leq L \|x - x^{*}\|_{\infty} \leq d, \text{ for } x \in N_{d}.  
		\end{align*}
		
		Therefore, $ g(\cdot): N_{d} \rightarrow N_{d} $, and by the contraction mapping theorem, we have that $ x^{a+1} = g(x^{a}) $ converges to $ x^{*} $ if the algorithm starts in the neighborhood $ N_{d} $. 
\end{proof}

\section{Appendix C}
\begin{figure}
	\begin{center}
		\begin{tabular}{c}
			\includegraphics[width=16cm]{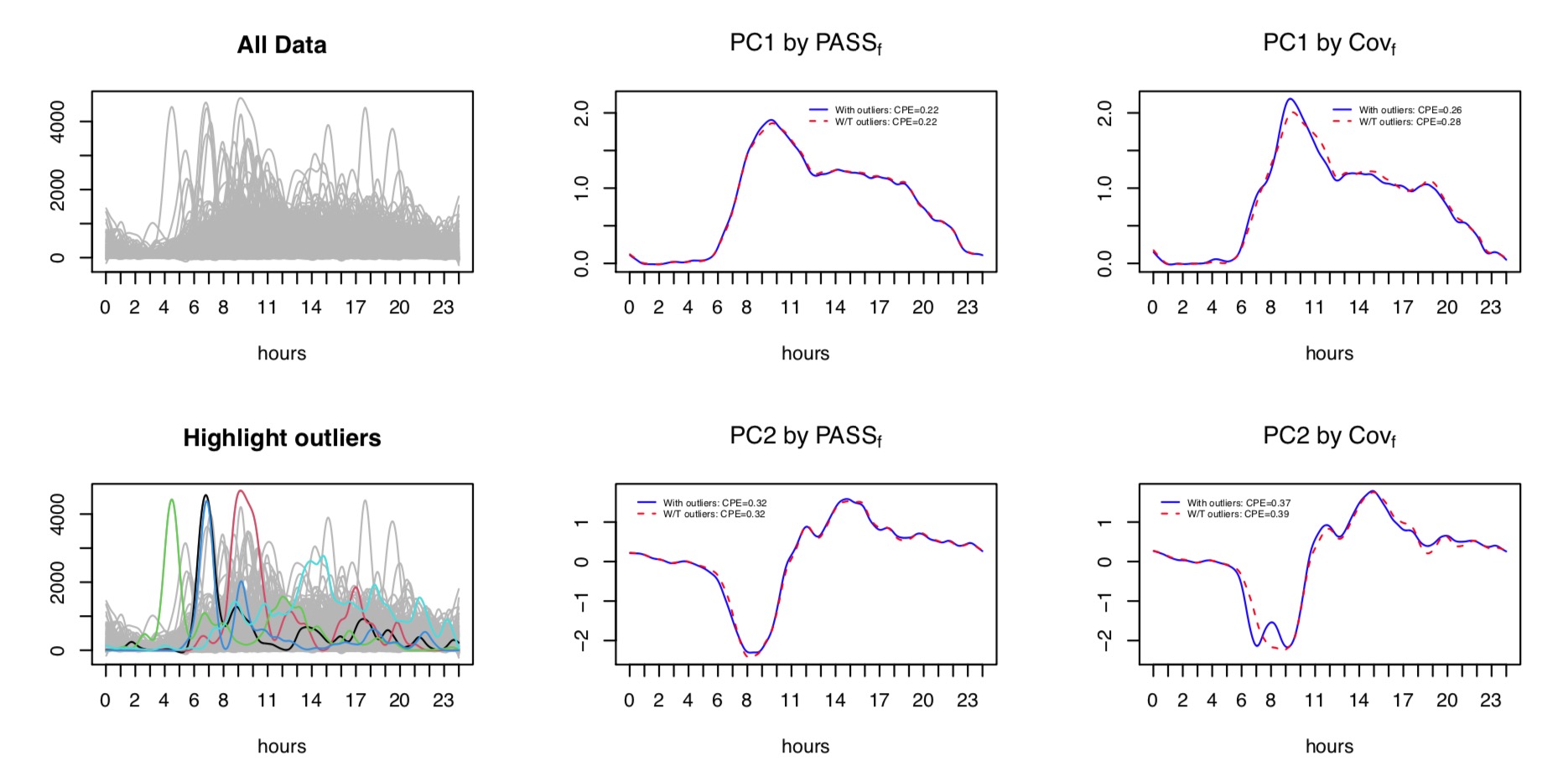}
		\end{tabular}
	\end{center}
	\caption{500 Accelerometery data: Estimated eigenfunctions and cumulative percentage of variation (CPE) by PASS$_f$ and Cov$_f$ with outliers included v.s. outliers removed. Outliers identified by directional quantile with R package fdaoutlier \citet{Ojo2020}} \label{app efun 1}
	\vspace{-0.1in}
\end{figure}

\begin{figure}
	\begin{center}
		\begin{tabular}{c}
			\includegraphics[width=16cm]{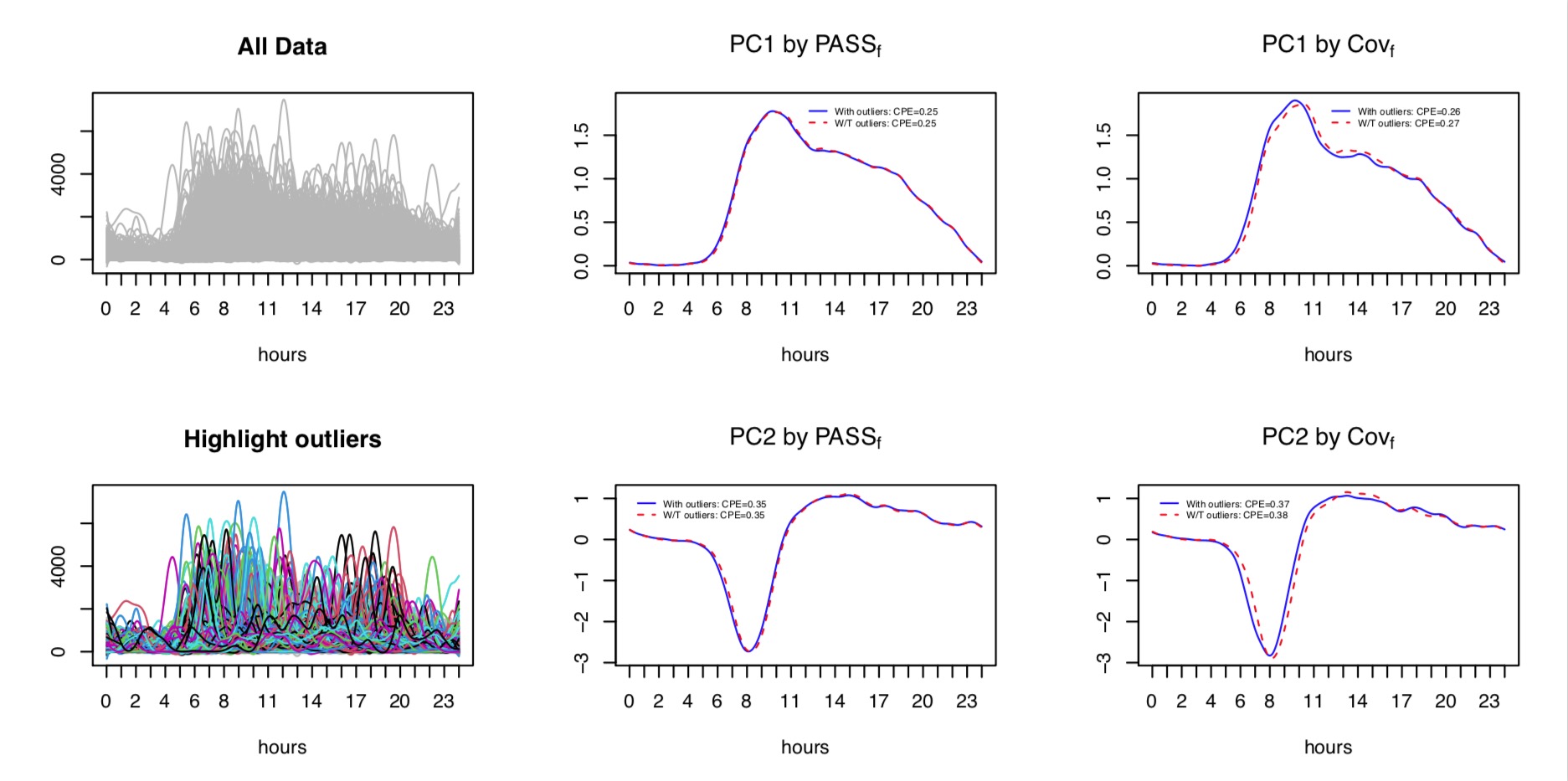}
		\end{tabular}
	\end{center}
	\caption{6389 Accelerometery data: Estimated eigenfunctions and cumulative percentage of variation (CPE) by PASS$_f$ and Cov$_f$ with outliers included v.s. outliers removed. Outliers identified by directional quantile with R package fdaoutlier \citet{Ojo2020}} \label{app efun 2} 
	\vspace{-0.1in}
\end{figure}

\begin{figure}
	\begin{center}
		\begin{tabular}{c}
			\includegraphics[width=16cm]{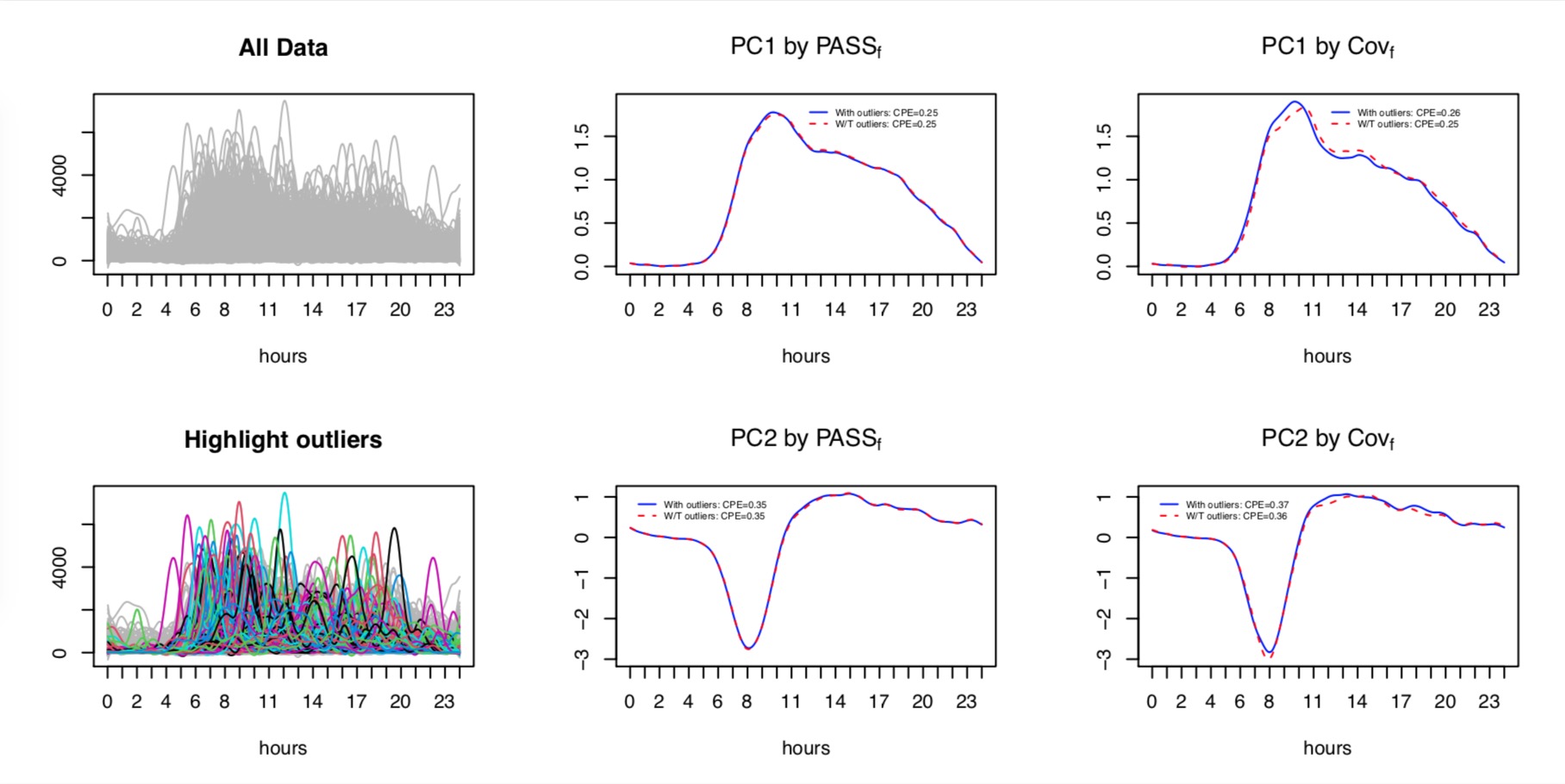}
		\end{tabular}
	\end{center}
	\caption{6389 Accelerometery data: Estimated eigenfunctions and cumulative percentage of variation (CPE) by PASS$_f$ and Cov$_f$ with outliers included v.s. outliers removed. Outliers are identified by their L$^2$ norm with a cutoff value 800.} \label{app efun 3}
	\vspace{-0.1in}
\end{figure}

\begin{figure}
	\centering
	\includegraphics[scale=0.54]{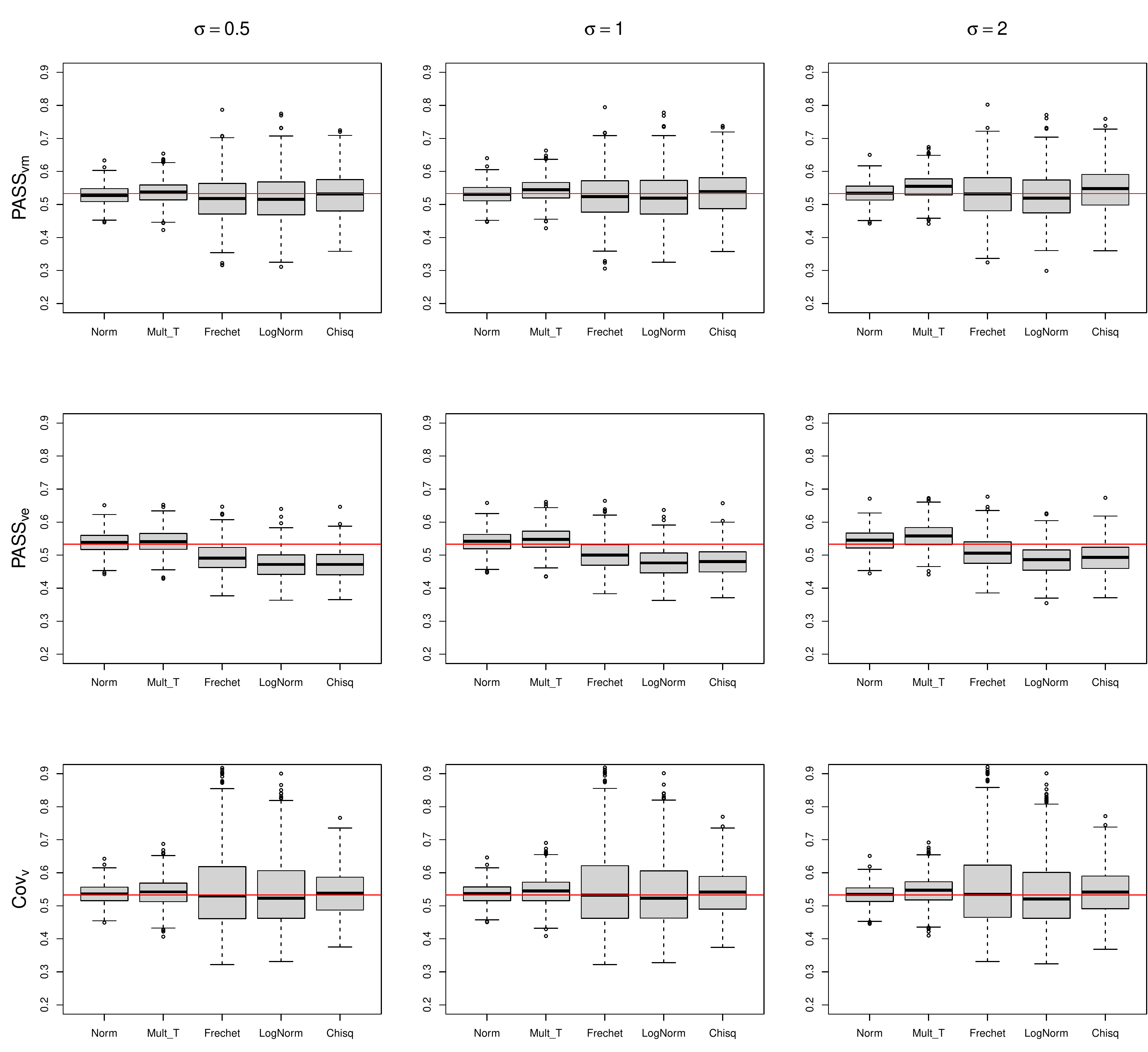}
	\caption{\small Boxplots of the estimated percentage of variance explained by the first eigenvalue when sample curves are observed with mean $0$, standard deviation $\sigma$ normal random noise. The red horizontal line indicates the true value. Sample size is 200. The first two rows are the PASS based methods, where PASS$_{vm}$ uses Algorithm 1 and PASS$_{ve}$ uses the elliptical distribution induced integral representation based algorithm in \cite{Durre2016a}. The third row is the covariance function based method (Cov$_{v}$). All three methods are applied to pre-smoothed curves. Columns correspond to $\sigma = 0.5,\ 1,\ 2$ respectively. No additional outliers are added.}
	\label{simulation fig: eigenvalue with noise}
\end{figure}

This section contains application eigenfunction estimation results (Figures \ref{app efun 1}, \ref{app efun 2} and \ref{app efun 3}). Simulation eigenratio estimations results when data are sampled with noise is also included (Figure \ref{simulation fig: eigenvalue with noise}).

\end{document}